\documentclass[aps,prb,reprint,superscriptaddress]{revtex4-2}
\usepackage{graphicx} 
\usepackage{tikz}
\usepackage{geometry}
\usepackage{physics}
\usepackage{placeins}
\usepackage{amsmath}
\usepackage{amssymb}

\newcommand{\emptycircle}{
  \begin{tikzpicture}[baseline=-0.6ex]
    \draw (0,0) circle (0.1);
  \end{tikzpicture}
}
\newcommand{\fullcircle}{
  \begin{tikzpicture}[baseline=-0.6ex]
     \filldraw[fill=black] (0,0) circle (0.1);
  \end{tikzpicture}
}

\begin{document}
\title{The quantum Kibble-Zurek mechanism: the role of boundary conditions, endpoints and kink types}

\author{Jose Soto-Garcia}

\affiliation{Kavli Institute of Nanoscience, Delft University of Technology, Lorentzweg 1, 2628 CJ Delft, The Netherlands}
\author{Natalia Chepiga}
\affiliation{Kavli Institute of Nanoscience, Delft University of Technology, Lorentzweg 1, 2628 CJ Delft, The Netherlands}

\begin{abstract}
Quantum phase transitions are characterised by the universal scaling laws in the critical region surrounding the transitions. This universality is also manifested in the critical real-time dynamics through the quantum Kibble-Zurek mechanism. In recent experiments on a Rydberg atom  quantum simulator, the Kibble-Zurek mechanism has been used to probe the nature of quantum phase transitions.
 In this paper we analyze the caveats associated with this method and develop strategies to improve its accuracy. Focusing on two minimal models---transverse-field Ising and quantum three-state Potts, both in one dimension---we study the effect of boundary conditions, the location of the endpoints and some subtleties in the definition of the kink operators. 
 In particular, we show that the critical scaling of the most intuitive types of kinks is extremely sensitive to the correct choice of endpoint, while more advanced types of kinks exhibit remarkably robust universal scaling. Furthermore, we show that when kinks are tracked over the entire chain, fixed boundary conditions improve the accuracy of the scaling. Surprisingly, the Kibble-Zurek critical scaling appears to be equally accurate whether the fixed boundary conditions are chosen to be symmetric or anti-symmetric.  We also show that the density of kinks extracted in the central part of long chains obeys the predicted universal scaling for all types of boundary conditions. Finally, we test our kink definition for the Ising transition on the period-2 phase of the Rydberg model and show that it is more robust against the end point than the standard definition.
 \end{abstract}


\maketitle

\section{Introduction}
\label{sec:intro}
Understanding the nature of quantum phase transitions is a central challenge in condensed matter physics. Unlike classical  transitions, which are driven by thermal fluctuations, quantum phase transitions (QPTs) appear due to quantum fluctuations, offering a unique perspective on the fundamental mechanisms of critical phenomena \cite{sachdev2011quantum}. 

Traditionally, the critical exponents characterizing QPTs have been investigated in equilibrium \cite{cardy1996scaling}. However, the discovery of the  Kibble-Zurek (KZ) mechanism \cite{zurek2005dynamics,polkovnikov2005, uhlmann2007vortex, dziarmaga2010dynamics,keesling2019quantum} has opened new pathways to explore the interplay between non-equilibrium dynamics and critical phenomena. Near the critical point of a second-order phase transition, both the correlation length $\xi$ and healing time $\tau$ diverge with the distance to the transition $g_c$  following a power law scaling\cite{dziarmaga2010dynamics}:
\begin{align}
    \xi & \sim \abs{g - g_c}^{-\nu},\\
    \tau & \sim \abs{g - g_c}^{-\nu z}.
\end{align}

When a system is driven with a constant sweep rate $s$ from a disordered phase to an ordered one crossing a second-order phase transition, its dynamics become non-adiabatic when the temporal distance to the critical point matches the system's healing time. This loss of adiabaticity leads to the formation of topological defects---kinks $n_k$ \cite{uhlmann2010n, uhlmann2010system}. Sufficiently far from the transition, the density of kinks scales as a power law with the sweep rate \cite{del2014universality}: 
\begin{equation}
    n_k \sim s^{(D - d)\mu},
\end{equation}
where $D$ and $d$ denote the spatial dimensionalities of the system and the defects, respectively, and $\mu$ is Kibble-Zurek critical exponent that can be expressed as a function of the critical exponent $\nu$ and the dynamical critical exponent $z$:

\begin{equation}
    \mu = \frac{\nu}{1+\nu z}.    
\end{equation} 
In one dimension, these kinks are quasiparticle excitations that act as domain walls between ordered regions \cite{uhlmann2010n}.

Originally proposed by Kibble to explain the formation of topological defects in the early universe \cite{kibble1976topology, kibble1980some}, the symmetry-breaking framework was later extended by Zurek to describe classical \cite{zurek1985cosmological} and quantum \cite{zurek2005dynamics, polkovnikov2005, dziarmaga2005dynamics} phase transitions in condensed matter systems. In its original formulation, the Kibble–Zurek mechanism (KZM) postulated that the system effectively freezes at the onset of the non-adiabatic regime, when the relaxation time exceeds the inverse rate of change of the control parameter \cite{zurek2005dynamics}. However, subsequent studies have shown that this freeze-out is only approximate. The system continues to evolve, with correlations spreading via the propagation of quasiparticle excitations. This extended dynamical behavior is captured by the sonic-horizon hypothesis, which refines the original KZM by accounting for the post-critical spreading of correlations \cite{kolodrubetz2012nonequilibrium, chandran2012kibble, francuz2016space, sadhukhan2020sonic}.

Experimental progress in platforms of Rydberg atoms \cite{bernien2017probing} have brought renewed attention to the experimental study of quantum many-body systems, and in particular, to non-equilibrium critical phenomena\cite{keesling2019quantum}, spurring extensive numerical \cite{laguna1997density, saito2007kibble, de2010spontaneous, del2010structural, jaschke2017critical, soto2024resolving} and experimental \cite{weiler2008spontaneous, griffin2012multiferroics, lamporesi2013spontaneous, mielenz2013trapping, pyka2013topological, ulm2013observation, beugnon2017exploring, ko2019kibble, cui2020experimentally, bando2020probing, lee2024universal} studies of the quantum KZ mechanism. However, although these numerical and experimental studies typically report power-law scaling, the critical exponents they extract often deviate from those predicted by the corresponding universality class.  Such discrepancies are believed to arise from competing effects inherent to the KZ protocol, which can obscure the underlying universal behavior \cite{rigol2008thermalization, del2022locality, samajdar2024quantum}.

The remainder of this article is organized as follows. In Section \ref{sec:Models}, we introduce the two one-dimensional (1D) models studied in this work: the transverse field Ising model and the quantum 3-state Potts model. Section \ref{sec:Methods} provides a concise review of the Tensor Network algorithms used in our simulations, along with relevant technical details. In Section \ref{sec:SS}, we investigate the influence of system size on the Kibble–Zurek mechanism. Section \ref{sec:Kinks} presents an alternative approach to counting kinks by focusing exclusively on isolated kinks, and we show that this method is more robust to changes in the final point of the sweep through the transition. In Section \ref{sec:Boundaries}, we compare various boundary conditions, demonstrating that fixing the boundaries—either on the same or in different directions—yields more accurate results than leaving them free. Section~\ref{sec:Rydberg} applies our alternative kink definition to the Ising transition in the Rydberg atoms model. Finally, Section \ref{sec:Conclusion} summarizes our findings and offers concluding remarks.

\section{The Models}
\label{sec:Models}
We explore the role of boundary conditions and the effectiveness of various definitions of kinks in two well established microscopic quantum models: The antiferromagnetic transverse field Ising model and the quantum 3-state Potts model in 1D.
\subsection{Transverse Field Ising Model}
The transverse field Ising model (for simplicity we will refer to it as Ising model) is defined by the following microscopic Hamiltonian:
\begin{equation}
    H = J\sum_{i=1}^{L-1}\sigma^z_i\sigma^z_{i+1} - h\sum_{i=1}^L\sigma^x_i,
\end{equation}
where $\sigma^{z,x}$ represents the Pauli matrices, $J$ is the nearest-neighbor Ising coupling, and $h$ is an external magnetic field in the transverse direction. The model is ferromagnetic when $J<0$ and antiferromagnetic when $J>0$. In this paper, we focus on the latter case. The location of the critical point $h/J=1$ is known exactly, along with all relevant critical exponents: $z=1$, $\nu=1$, $\mu=1/2$\cite{difrancesco}. To control the boundary conditions we also included a magnetic field in the first and last lattice sites along the $z-$direction.
\begin{equation}
    H = H_0 -  (h_{z1}\sigma^z_1 + h_{zL}\sigma^z_L).
\end{equation}
For the antiferromagnetic Ising model ($J>0$), the standard number of kinks operator is defined as:
\begin{equation}
    n_k = \frac{1}{2}\sum_{i=1}^{L-1}\left(1 + \left<\sigma^z_i\sigma^z_{i+1}\right> \right),
\label{eq:kink_ising}
\end{equation}
which assigns 1 for each pair of aligned neighboring spins (i.e., a ‘kink’ in the otherwise alternating antiferromagnetic configuration).

\subsection{1D Quantum 3-state Potts Model}
The ferromagnetic quantum 3-states Potts model, to which we will refer as Potts model for simplicity, is a generalization of the ferromagnetic Ising model for the local Hilbert space of dimension three. The microscopic Hamiltonian is given by:
\begin{equation}
    H = -J\sum_{i = 1}^{L-1}\sum_{a = 1}^3 P_i^aP_{i+1}^a - h\sum_{i=1}^LP_i,
\end{equation}
where $P_i^a = \ket{a}\bra{a}_i - \frac{1}{3} \mathbb{I}$ is a traceless Hermitian operator derived from the projector onto the state $\ket{a}$ at site $i$, and $P_i = \ket{\lambda_0}\bra{\lambda_0} - \frac{1}{3} \mathbb{I}$ is the traceless Hermitian operator associated with projection along the state $\ket{\lambda_0} = \frac{1}{\sqrt{3}} \sum_a \ket{a}$.
 The first term represents a ferromagnetic interaction, and the second term represents a generalized transverse field. The critical point is located at $h/J = 1$ and has critical exponents $\nu = 5/6$ and $z = 1$, and thus $\mu \approx 0.454$\cite{difrancesco}. Similarly to the Ising model, boundaries can be controlled by adding an external longitudinal field:
\begin{equation}
    H = H_0 - \sum_{a=1}^{3} \left( h_{a1}P^a_1 + h_{aL}P^a_L \right). 
\end{equation}
The kink operator has the form:
\begin{equation}
    n_k = \sum_{i=1}^{L-1}\left(1 - \left<\sum_a (\ket{a}\bra{a})_i \cdot(\ket{a}\bra{a})_{i+1}\right>\right),
\label{eq:kink_potts}
\end{equation}
where  $(\ket{a}\bra{a})_i$ is the projector onto state $\ket{a}$ at site i. Thus $n_k$ counts a kink whenever two neighboring sites differ. For simplicity, we will refer to the three possible directions as $A$, $B$ and $C$.

\section{Methods}
\label{sec:Methods}
\subsection{Ground State Calculations}
The initial state defined at time $t=0$ is a ground state at a given starting point in the disordered phase sufficiently far from the transition. To ensure this point was in the adiabatic regime we set the starting point $h_0=10\cdot s^{1/(1+z\nu)}$. The ground state was determined using the density matrix renormalization group (DMRG) algorithm\cite{white1992density, schollwock2011density} on the matrix product state (MPS) formalism. Singular values were kept above $10^{-6}$, and maximal bond dimension was restricted to $D = 300$. Convergence of the ground state was assumed when the energy difference between two successive sweeps, including an increase in the bond dimension, divided by the system size was not exceeding $10^{-8}$. 
\subsection{Simulation of dynamics}
Simulations of the time evolution for the Ising model with open boundary conditions and Potts model were conducted using a second-order time-evolving block decimation (TEBD) algorithm \cite{vidal2004efficient, verstraete2004matrix, paeckel2019time}. In both cases, the quench protocol is conducted by keeping the coupling $J$ constant, while varying the transverse field linearly with time:
\begin{equation}
    h/J = -s\cdot t + h_0.
\end{equation}
As noted in the Introduction, we propose an alternative definition of kinks in this work; however, unless otherwise specified, kink counting follows the standard definitions given in Eqs.~(\ref{eq:kink_ising}) and (\ref{eq:kink_potts}) for the Ising and Potts models, respectively.

For the Rydberg model with $1/r^6$ van der Waals interactions, time evolution was simulated using the Time-Dependent Variational Principle (TDVP) \cite{haegeman2011time, haegeman2016unifying}. The long-range interactions were approximated using a sum of 7 exponentials\cite{pirvu2010matrix, schollwock2011density}, i.e., $1/r^6 = \sum_{i=1}^{11} c_i\lambda_i^r$. The coefficients $c_i$ and exponents $\lambda_i$ were determined by minimizing the cost function defined as:
\begin{equation}
\sum_{r=1}^L\abs{\frac{1}{r^6}- \sum_{i=1}^{7}c_i\lambda_i^r}.
\end{equation}
This optimization process followed the method described in \cite{pirvu2010matrix}. The maximum error in this approximation was $\sim 10^{-10}$, with a cost function $\sim 1.3 \times 10^{-19}$.

\subsection{Estimation of the uncertainty on the fitting}
To quantify the uncertainty in the KZ critical exponent $\mu$, we analyzed the stability of the fit with respect to the choice of data points. From the full set of sweep rates used for the fitting, we constructed sliding windows of four consecutive points and extracted an effective exponent  $\mu_{\textnormal{eff}}$ from each window. The total fitting error for $\mu$ was defined as the larger of the following two contributions:
\begin{itemize}
    \item the maximum deviation between the fitted value of $\mu$ and the set of effective exponents $\mu_{\textnormal{eff}}$;
    \item the $95\%$ confidence interval of the fit, obtained as $\pm1.96$ times the standard error of the fitted slope.
\end{itemize}
Unless explicitly stated, all data points shown in the figures were used for the fitting.

\section{System Size}
\label{sec:SS}
\begin{figure}[!htb]
    \centering
    \includegraphics[width=0.95\linewidth]{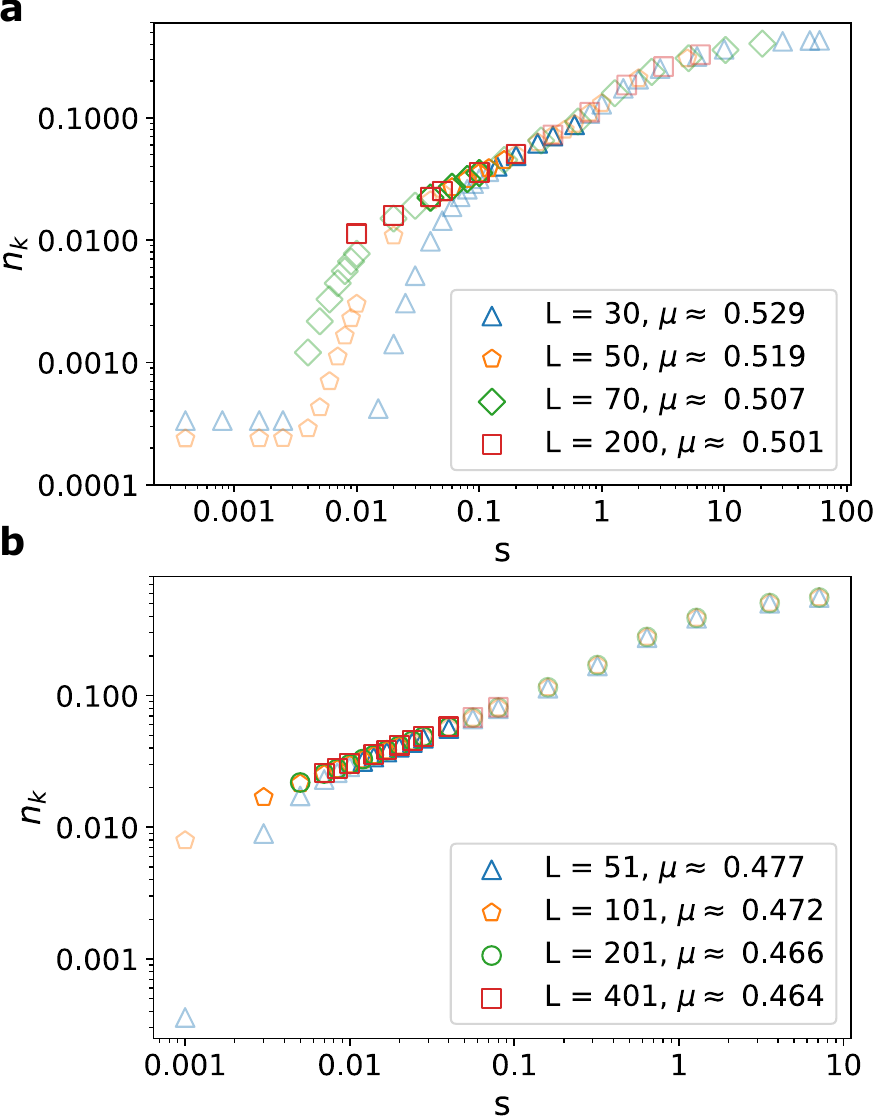}
    \caption{Kibble–Zurek scaling of the kink density $n_k$ as a function of the sweep rate $s$ for different system sizes in (a) the Ising and (b) Potts models. In both panels, three distinct dynamical regimes are visible. At slow sweep rates, the system evolves adiabatically, and the kink density remains near zero. As the sweep rate increases, an intermediate regime emerges in which $n_k$ exhibits universal power-law scaling with $s$. At high sweep rates, the dynamics become fully non-adiabatic, and the kink density saturates to a constant value.
    (a) For system sizes $L \geq 50$, the numerical results are in excellent agreement with the theoretical prediction $\mu = 0.5$.
    (b) A small finite-size effect leads to a slight overestimation of the critical exponent $\mu$ in smaller systems. For $L = 401$, the extracted exponent agrees with the theoretical value $\mu = 5/11$ within $2\%$. Data points not included in the KZ fits are shown in pale. All simulations were performed with a maximum bond dimension of $D = 300$.}
    \label{fig:systemsize}
\end{figure}

We first investigated the role of finite-size effects in the Kibble-Zurek (KZ) mechanism for both models. Fig.~\ref{fig:systemsize}(a) summarizes our findings for the Ising model, where we explored a wide range of sweep rates $s$ to capture the full spectrum of dynamical behaviors. Three distinct regimes emerge clearly. For slow sweeps, the correlation length becomes comparable to the system size, and the system evolves adiabatically throughout the process. In this regime, the kink density remains near zero and is largely independent of the sweep rate \cite{dziarmaga2005dynamics, dziarmaga2010dynamics}. For fast sweeps, the evolution is fully non-adiabatic, and the kink density saturates at a maximum value determined by the final point of the quench \cite{xia2021kibble, zeng2023universal}. Between these two extremes lies the KZ scaling regime, characterized by a power-law dependence of the kink density on the sweep rate.\footnote{ At the boundaries between the KZ regime and the other two regimes, we observe a transient behavior in which the kinks density increases more rapidly with the sweep rate than the KZ scaling would predict. The transient regime between the fast-quench and the KZ regime is the pre-saturated regime, which occurs when the initial point is in the adiabatic regime, but the ending point is inside the non-adiabatic regime  \cite{kou2023varying}. The size of both transient regimes is model dependent.}

We estimated the optimal fitting window by scanning over all contiguous subsets of the data containing at least a minimum of 6 data points (with no restriction on the maximum). For each candidate region, we performed a linear fit and evaluated its associated uncertainty. The region yielding the smallest fitting uncertainty was selected as the optimal scaling window, and the corresponding region was used in all subsequent analyses.

In both models, increasing the system size has a noticeable but moderate impact on the extracted Kibble–Zurek (KZ) scaling exponent $\mu$, as shown in Fig.~\ref{fig:systemsize}. For the Ising model [Fig.~\ref{fig:systemsize}(b)], this dependence is very weak: results for smaller systems can be brought closer to the theoretical prediction $\mu = 0.5$ by reducing the number of data points included in the fit. As the system size increases, the threshold sweep rate below which the scaling becomes linear shifts to smaller values, thereby extending the interval over which KZ scaling can be reliably observed. The extracted exponents are $\mu = 0.54 \pm 0.04$ for $L = 30$, $\mu = 0.515 \pm 0.020$ for $L = 50$, $\mu = 0.512 \pm 0.005$ for $L = 70$, and $\mu = 0.5014 \pm 0.0016$ for $L = 200$.

For the 3-state Potts model [Fig.~\ref{fig:systemsize}(b)], reducing the number of data points does not improve the agreement with the predicted value at small system sizes. Instead, the slope of the scaling curve systematically approaches the theoretical value $\mu = 5/11$ from above as $L$ increases. The extracted exponents are $\mu = 0.477 \pm 0.004$ for $L = 51$, $\mu = 0.472 \pm 0.003$ for $L = 101$, $\mu = 0.466 \pm 0.004$ for $L = 201$, and $\mu = 0.464 \pm 0.004$ for $L = 401$. For $L = 401$, the value of $\mu$ agrees with the theoretical prediction to within 2\%.

\section{Types of kinks and the endpoint}
\label{sec:Kinks}
The standard experimental protocol for extracting the KZ exponent $\mu$ consists of driving the system from the disordered phase into the ordered phase and counting the number of kinks after the transition for various sweep rates. In our simulations, we evaluate the expectation value of the kink operator rather than sampling individual spin configurations.

As mentioned in the introduction, in a first approximation the system freezes upon entering the nonadiabatic regime, and the correlation length (which in 1D is directly related to the density of kinks) remains fixed thereafter. However, this picture is oversimplified: the state continues to evolve and correlations keep spreading even after crossing the critical point~\cite{kolodrubetz2012nonequilibrium, chandran2012kibble, francuz2016space, sadhukhan2020sonic}. In the Ising model, a more refined description shows that the KZ mechanism excites entangled kink pairs with opposite momenta, which propagate correlations throughout the system~\cite{kolodrubetz2012nonequilibrium, chandran2012kibble}. Owing to their fermionic character, kink quasiparticles exhibit strong short-range anticorrelations, which drastically suppress the probability of generating two kinks on adjacent bonds. As a result, the occurrence of consecutive kinks produced by the KZ mechanism is essentially forbidden~\cite{Dziarmaga2022kink}.

Moreover, during the dynamics, quantum coarsening competes with the KZ mechanism, further increasing the correlation length \cite{samajdar2024quantum}. As a result, the correlation length continues to increase even after the system has returned to the adiabatic regime.

Finally, if the transverse field is not completely suppressed at the end of the ramp, the ground state is not a perfectly ordered state, but there is a certain probability of local spin flips. Consequently, not all detected kinks arise from the KZ mechanism itself. From perturbation theory, one can estimate the density of spin flips for a given value of the transverse field. For both the Ising and Potts model considered in this work, the ground state in the presence of a small transverse field $h$ takes, to first order in perturbation theory, the form
\begin{equation}
    \ket{\psi_0} \approx \ket{0} + \frac{h}{4J}\frac{1}{\sqrt{L}}\sum_i\ket{f_i} + O(h^2),
\end{equation}
where $\ket{0}$ is the ground state for $h=0$, and $\ket{f_i}$ is a state with a spin flip in position $i$. It is a standard practice in this type of protocol to remove the expectation value of the ground state to kink distribution after the quench to eliminate the influence of the endpoint~\cite{sen2010quenching, sharma2015one, hegde2015quench}. However, the protocol presented here has the advantage that it does not require precise knowledge of the endpoint. 

\begin{figure}[!htb]
    \centering
    \includegraphics[width=0.95\linewidth]{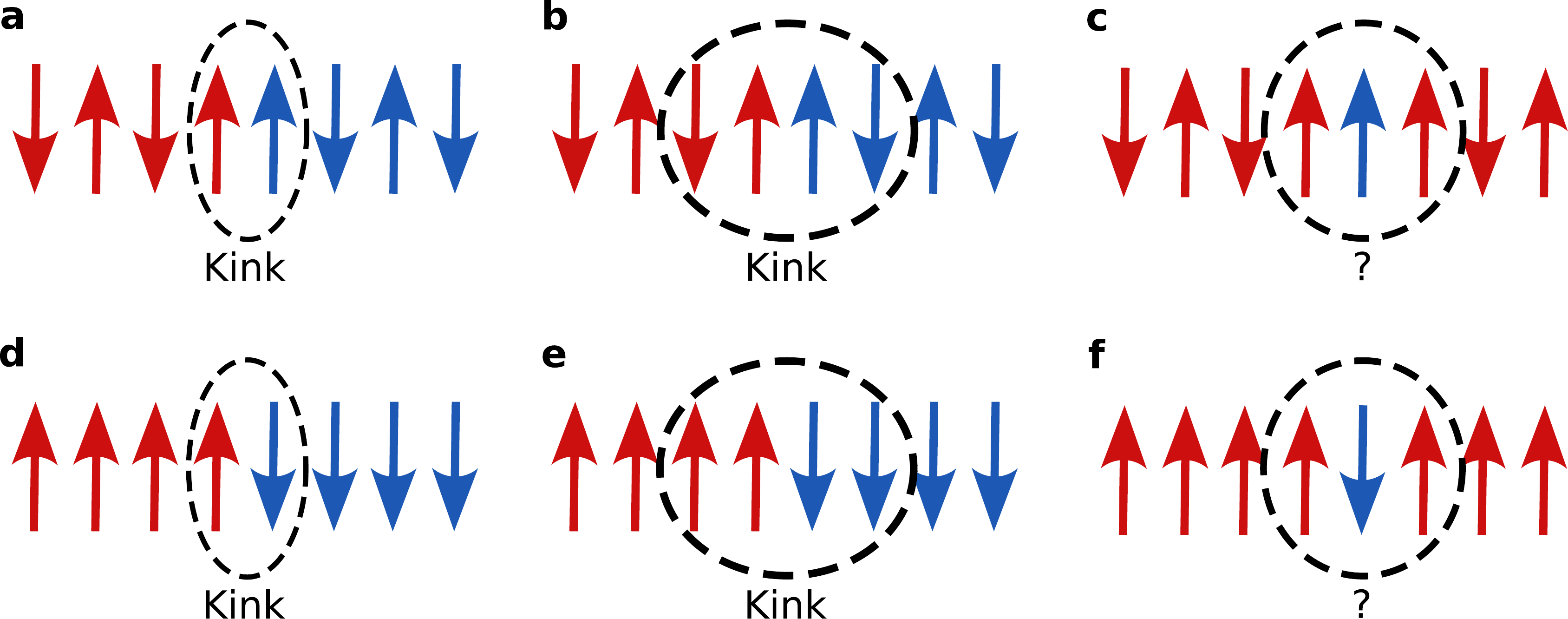}
    \caption{Kinks and domain walls in the Ising model with (a)-(c) antiferromagnetic and (d)-(e) ferromagnetic interactions. Red and blue colors are assigned to each of the two ground-states. (a),(d)In the standard definition \cite{dziarmaga2010dynamics}, a kink is counted whenever the expected ordering is violated. (b,e) In the "isolated kink" definition proposed in this work, kinks are identified over four consecutive spins, requiring that the domains on either side of the kink have a minimum length of two. (c,f) A single spin-flip is interpreted as a double kink under the standard definition, but is excluded in the isolated kink criterion, since it does not alter the nature of the surrounding domains. We argue that such isolated spin flips should not contribute to the kink count, as they do not correspond to genuine domain boundaries.}
    \label{fig:isingkinkoperator}
\end{figure}

In this section, we analyze two factors that affect the calculation of the KZ critical exponent $\mu$: the endpoint where the dynamics terminates and the choice of a kink operator. All results presented in this section were obtained with fixed boundary conditions, while the effect of various boundary conditions on the apparent KZ critical exponent will be discussed in detail in the next section.  We consider two types of kink definitions, illustrated in Fig.~\ref{fig:isingkinkoperator}, corresponding to the Ising chain with antiferromagnetic interactions (Figs.~\ref{fig:isingkinkoperator}(a)–\ref{fig:isingkinkoperator}(c)) and with ferromagnetic interactions (Figs.~\ref{fig:isingkinkoperator}(d)–\ref{fig:isingkinkoperator}(f)). In the standard definition, a kink is counted whenever the expected order is violated. For instance, 
in the ferromagnetic case, a kink is counted if any pair of neighboring spins are anti-aligned, while in the antiferromagnetic case, it is counted if they are aligned (see Fig.~\ref{fig:isingkinkoperator}(d) and (a) correspondingly). 

Alternatively, one can count only isolated kinks, thus excluding configurations in which more than two consecutive spins deviate from the expected order. In the antiferromagnetic case, this means disregarding regions where three or more neighboring spins are aligned instead of alternating: 
\begin{equation}
\begin{aligned}
n_k = \sum_i \left(
    \left\langle P_{\uparrow i} \otimes P_{\downarrow i+1} \otimes P_{\downarrow i+2} \otimes P_{\uparrow i+3} \right\rangle \right. \\
    \left. + \left\langle P_{\downarrow i} \otimes P_{\uparrow i+1} \otimes P_{\uparrow i+2} \otimes P_{\downarrow i+3} \right\rangle
\right)
\end{aligned}
\end{equation}
with $P_\uparrow$ and $P_\downarrow$ are the projectors for spin up and down respectively, as shown in Fig.~\ref{fig:isingkinkoperator}(b). In the ferromagnetic case, it entails excluding regions where three or more spins are anti-aligned:
\begin{equation}
\begin{aligned}
n_k = \sum_i \left(
    \left\langle P_{\uparrow i} \otimes P_{\uparrow i+1} \otimes P_{\downarrow i+2} \otimes P_{\downarrow i+3} \right\rangle \right. \\
    \left. + 
    \left\langle P_{\downarrow i} \otimes P_{\downarrow i+1} \otimes P_{\uparrow i+2} \otimes P_{\uparrow i+3} \right\rangle
\right)
\end{aligned}
\end{equation}
as depicted in Fig.~\ref{fig:isingkinkoperator}(e). Such scenarios are typical for a single-spin flip, as shown Fig.~\ref{fig:isingkinkoperator}(c) and (f). Under the conventional definition~\cite{dziarmaga2010dynamics}, such a single flip would count as two kinks. 

\begin{figure}[!htb]
    \centering
    \includegraphics[width=\linewidth]{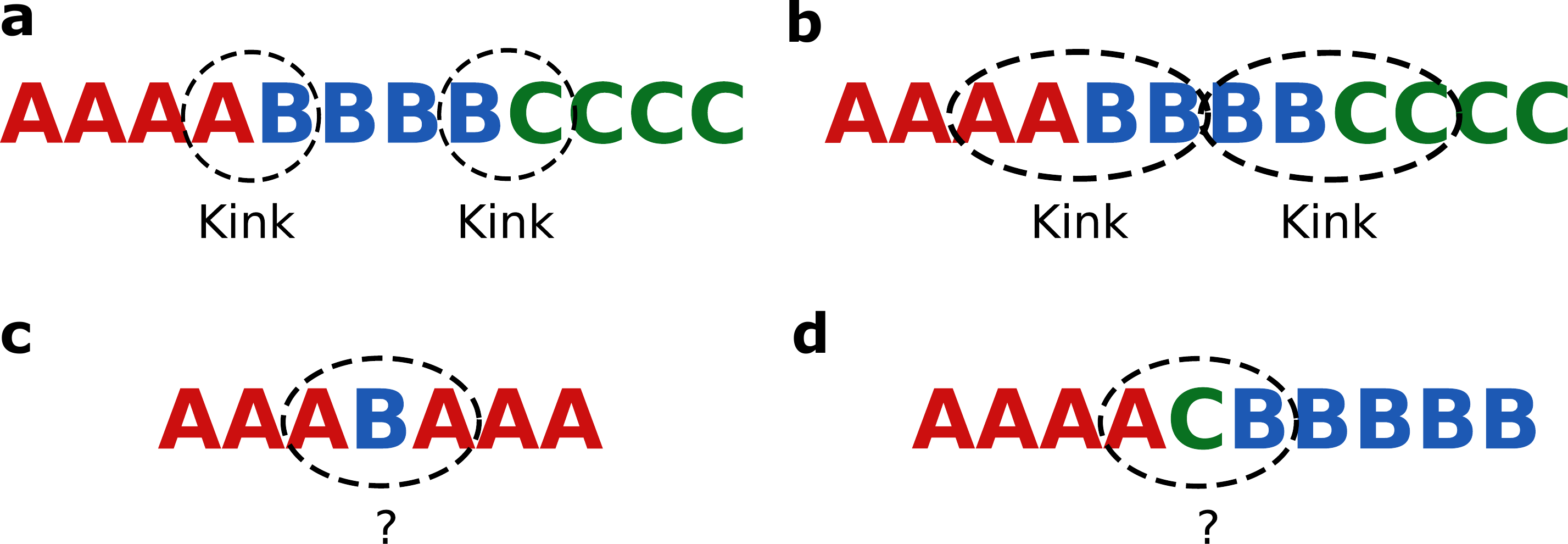}
    \caption{Domain wall characterization in the three-state Potts model. (a) In the standard definition, a kink is counted whenever two consecutive sites occupy different local states. (b) In the "isolated kink" definition proposed in this work, a kink is only counted when two differing spins are also aligned with their nearest neighbors, ensuring that the adjacent domains have a minimum length of two. (c) A single spin-flip is interpreted as a double kink under the standard definition, but is excluded in the isolated kink definition, as it does not alter the structure of the surrounding domain. (d) A spin-flip occurring at the boundary between two genuine domains is counted as a double kink in the standard approach. In contrast, the isolated kink definition counts it as a single kink, since only two true domains are present in the configuration.}
    \label{fig:pottskinkoperator}
\end{figure}

This logic of robust kinks can be extended beyond the trivial Ising model, as shown in Fig.~\ref{fig:pottskinkoperator}. However, for the Potts model, the definition of a robust kink is more subtle, since the spin flip might be located between two different domains, as shown in Fig.~\ref{fig:pottskinkoperator}(c)-(d). In this case, the advanced definition of kink discards spin flips inside a domain of the same type (Fig.~\ref{fig:pottskinkoperator}(c)), but treats a spin flip that bridges two different domains as a single kink (Fig.~\ref{fig:pottskinkoperator}(d)):
\begin{equation}
\begin{aligned}
n_k = 
\sum_i \left(
    \sum_{a \neq b}
        \left\langle 
            (\ket{a}\bra{a})_i
            (\ket{a}\bra{a})_{i+1}
            (\ket{b}\bra{b})_{i+2}
            (\ket{b}\bra{b})_{i+3}
        \right\rangle
        \right. \\
    \left.
    + 
    \sum_{\substack{a,b,c \\ a\neq b,\, b\neq c,\, a\neq c}}
        \left\langle
            (\ket{a}\bra{a})_i
            (\ket{b}\bra{b})_{i+1}
            (\ket{c}\bra{c})_{i+2}
        \right\rangle
\right)
\end{aligned}
\end{equation}

\begin{figure}
    \centering
    \includegraphics[width=\linewidth]{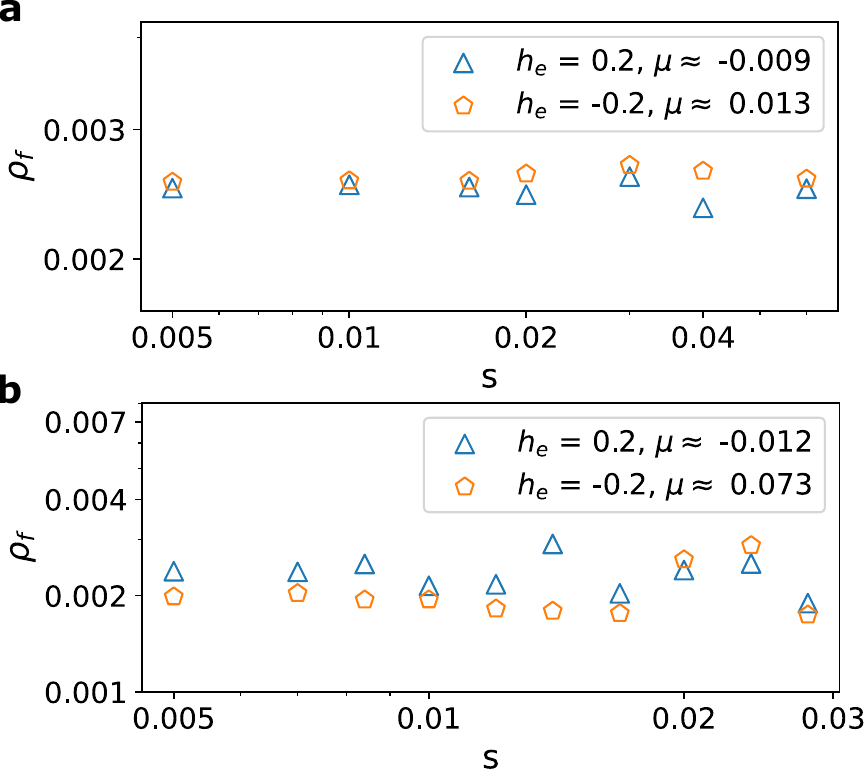}
    \caption{Density of spin flips for $\rho_f$ for (a) the Ising model and (b) the 3-state Potts model as a function of the sweep rate $s$ for two end points $h_e$ of the transverse field. In both cases, $\rho_f$ is independent of the sweep rate. In both cases, the system size is $L = 201$.}
    \label{fig:kzspinflip}
\end{figure}

We begin by examining how the density of single spin flips depends on the sweep rate $s$. In Fig.~\ref{fig:kzspinflip}, we report the single--flip density $\rho_f$ as a function of $s$ for two final values of the transverse field, $h_e = \pm 0.2$, for both the Ising model (Fig.~\ref{fig:kzspinflip}(a)) and the Potts model (Fig.~\ref{fig:kzspinflip}(b)). 

For the Ising model, we obtain slopes $\mu = -0.009 \pm 0.030$ for $h_e = 0.2$ and $\mu = 0.013 \pm 0.017$ for $h_e = -0.2$. For the Potts model, the corresponding values are $\mu = -0.01 \pm 0.19$ and $\mu = 0.07 \pm 0.22$. Within uncertainties, all these slopes are consistent with zero. 

\begin{figure}
    \centering
    \includegraphics[width=\linewidth]{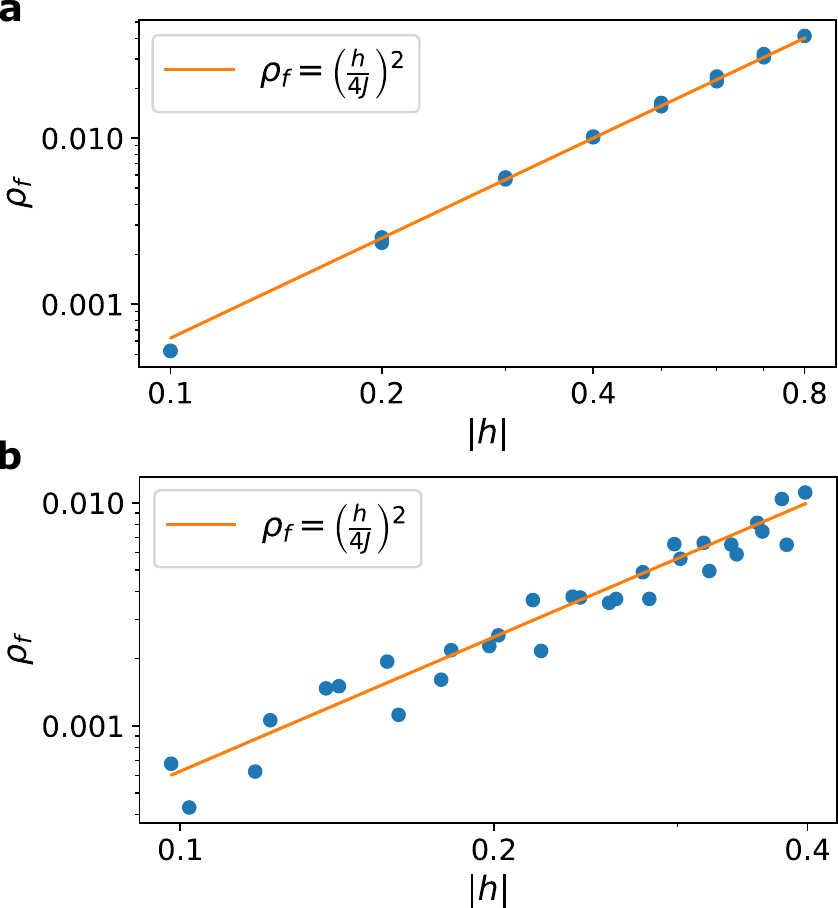}
   \caption{Density of single spin flips as a function of the absolute value of the transverse field $h$ during the quench for (a) the Ising model and (b) the 3-state Potts model, with $L = 201$ sites. The sweep rate is $s = 0.04$ for Ising and $s = 0.02$ for Potts. In both cases, the solid line shows the expected scaling of the spin-flip density obtained from first-order perturbation theory in the ground state.}
    \label{fig:evospinflip}
\end{figure}

In Fig.~\ref{fig:evospinflip} we show the density of spin flips generated during the quench for the Ising model with $s=0.04$ (Fig.~\ref{fig:evospinflip}(a)) and the 3-state Potts model with $s = 0.02$  (Fig.~\ref{fig:evospinflip}(b)) for $L = 201$ sites. In both cases, the resulting kink density closely follows the perturbative prediction for the ground state, confirming that the observed spin flips are consistent with the expected field-induced defect background.
\begin{equation}
    \frac{\bra{\psi_0}\hat f\ket{\psi_0}}{\braket{\psi_0}}\approx \left(\frac{h}{4J}\right)^2,
\end{equation}
confirming that the density of spin flips is independent of the sweep rate.

\begin{figure}[!htb]
    \centering
    \includegraphics[width=0.95\linewidth]{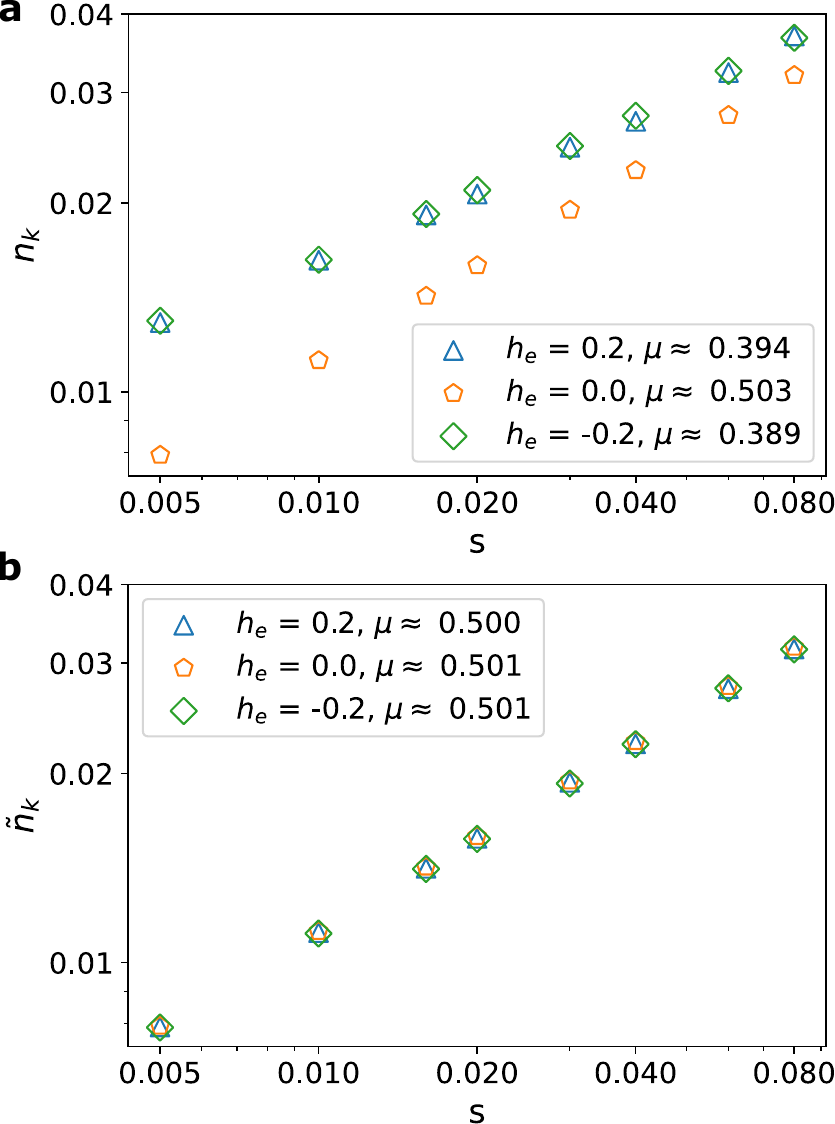}
    \caption{Scaling of the density of kinks $n_k$ with the sweep rate $s$ for the Ising model for various final values of the transverse field $h_e$ using (a) the standard kinks operator, (b) a kink operator that only counts isolated domain walls (i.e. excludes local spin flips). The Kibble-Zurek exponent $\mu$ is extracted from the slope of the power law scaling using the whole set of data points depicted in the figure. (a) Numerically extracted $\mu$ using the standard kink operator agrees with the theory prediction $\mu=0.5$ only for $h_e = 0$, while the value of $\mu$ extracted with other endpoints is significantly underestimated. (b) The scaling of isolated kinks provides accurate numerical estimates of the Kibble-Zurek critical exponent $\mu$ for a wide range of final endpoints. In both cases, the system size is $L = 201$.}
    \label{fig:tfimkinkstype}
\end{figure}

Fig.~\ref{fig:tfimkinkstype}(a) shows the critical scaling of the density of standard kinks with a sweep rate for three different endpoints: $h_e = 0$ and $h_e =\pm 0.2$. The extracted KZ critical exponent is $\mu = 0.5032\pm0.0012$ for $h_e = 0$, and $\mu = 0.39\pm0.05$ for $h_e = \pm0.2$. While the KZ exponent extracted for the trajectory terminating at $h_e = 0$ is in excellent agreement with the theory prediction for Ising criticality $\mu=0.5$, critical exponents extracted for other endpoints deviate significantly. If instead we use an advanced definition of kinks where a single-spin flip is explicitly excluded from the kink counting, extremely accurate values of KZ exponents have been extracted from time-evolution to all three endpoints, as demonstrated in Fig.~\ref{fig:tfimkinkstype}(b). In this case, the extracted critical exponent is $\mu = 0.5011\pm0.0011$ for $h_e = 0.2$, $\mu = 0.4997\pm0.0020$ for $h_e = 0.2$, and  $\mu = 0.5009\pm0.0019$ for $h_e = -0.2$. This means that local impurities do not play a role in the Kibble-Zurek dynamics and must generally be filtered out by a proper definition of the kink operators. Alternatively, one can terminate the dynamical evolution at the point where the quantum term responsible for the local spin flip is absent, and such impurities do not appear.

\begin{figure}[!htb]
    \centering
    \includegraphics[width=0.95\linewidth]{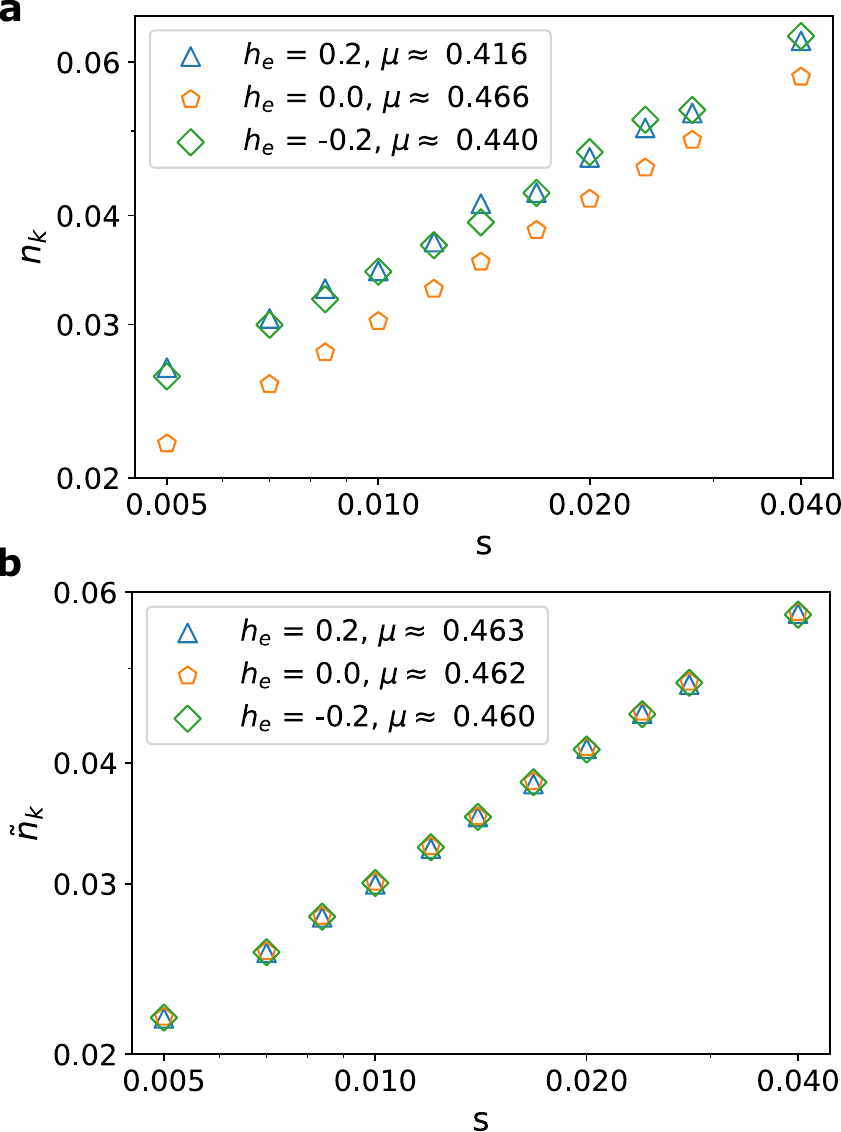}
    \caption{Scaling of the density of kinks $n_k$ with the sweep rate $s$ for the Potts model for various final values of the transverse field using (a) the standard kinks operator, (b) a kink operator that only counts isolated domain walls. The Kibble-Zurek exponent $\mu$ is extracted from the slope of the power law scaling using all data points observed in the figure. With the simple kink operator, only $\mu$ extracted for the trajectory that ends at $h_e = 0$ is reasonably close (within 3$\%$) to the theoretical value $\mu \approx 0.454$  for Potts, while other endpoints significantly reduce the slope. When using the isolated kink operator, numerically extracted value agree within $2\%$ with the theory prediction for a wide range of endpoints. In both presented cases the system size is $L = 201$.}
    \label{fig:pottskinkstype}
\end{figure}

This method of counting kinks in the Potts model produces results comparable to those of the Ising model, as shown in Fig.~\ref{fig:pottskinkstype}. The extracted KZ critical exponent for the standard kink operator (Fig.~\ref{fig:pottskinkstype}(a)) is $\mu = 0.466\pm0.004$ for $h_e = 0$, $\mu = 0.42\pm0.05$ for $h_e = 0.2$ and $\mu = 0.44\pm0.08$ for $h_e = -0.2$. The Kibble-Zurek exponent $\mu$ is closest to its theoretical value when the sweep ends at zero transverse field, $h_e = 0$, where the dynamical term vanishes.  For the advanced kink definition (Fig.~\ref{fig:pottskinkstype}(b)), the extracted KZ exponent is  $\mu = 0.4625\pm0.0022$ for $h_e = 0$, $\mu = 0.463\pm0.006$ for $h_e = 0.2$ and $\mu = 0.460\pm0.007$ for $h_e = -0.2$. This operator is more robust with respect to the ending point, causing the three curves to overlap and yielding a more accurate $\mu$, with much smaller uncertainty on the fitting. In all cases, the density of kinks scales as a power law dependence on the sweep rate.

\begin{figure}[!htb]
    \centering
    \includegraphics[width=0.95\linewidth]{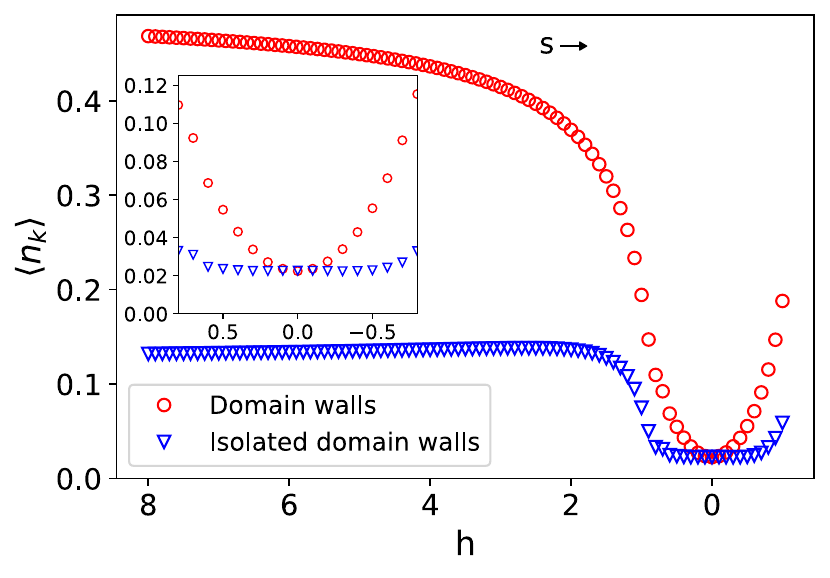}
   \caption{Evolution of the kink density $n_k$ during a constant-rate quench $s = 0.04$ in the finite size Ising model with $L = 201$, comparing the standard kink-counting method (red circles) to counting only isolated kinks (blue triangles), where isolated means that single spin flips are not counted. Deep in the disordered phase ($h \gg 1$), the density of kinks reaches its maximum ($n_k = 0.5$ for the standard definition and $n_k = 0.125$ for isolated kinks). As the system crosses the phase transition at $h = 1$, the emergence of large ordered regions leads to a sudden drop in the kink density. At $h=0$, both counting methods yield the same minimal density, since any remaining kinks are solely due to domain wall creation as described by the KZ mechanism. Notably, while isolated kinks maintain their minimal density over a wide range of $h$, the standard kink definition only attains this minimal value at $h=0$. The inset zooms into the region around $h=0$ for better clarity.}
    \label{fig:kinkste}
\end{figure}

Finally, let us compare the evolution of the overall density of standard and isolated kinks in the whole quench protocol. The results for the Ising model are presented in Fig.~\ref{fig:kinkste}.
Let us carefully explain this figure. When $h \rightarrow \infty$, the ground state is $\ket{...+++++...}$. The density of kinks in this maximally disordered state is $n_k = 0.5$ while isolated kinks is $n_k = 0.125$. This number can be easily obtained by counting the number of domain walls or isolated domain walls respectively in the perfectly disordered state (i.e. number of aligned pairs of spins or isolated aligned pairs of spins in the antiferromagnetic Ising model). As we approach the phase transition at $h = 1$, but still far from the non-adiabatic regime, the average number of kinks decreases, while the density of isolated kinks remains constant. Around the phase transition, the density of kinks drops sharply as ordered domains form, separated by a small number of domain walls. The density of standard kinks reaches a minimum at $h = 0$, where quantum fluctuations are solely driven by the Kibble-Zurek mechanism, and increases for $h > 0$. In contrast, the density of isolated kinks attains its minimum before $h = 0$ and remains nearly constant over a wide range of $h$ values on both sides of the transition. This indicates that the isolated kink operator is robust against random quantum fluctuations and selectively detects domain walls generated by the KZ mechanism.

\section{Boundary Conditions}
\label{sec:Boundaries}
The critical properties of a quantum system depend not only on the bulk Hamiltonian but also on its boundary conditions, and the same holds true for its dynamical behavior. In equilibrium boundary conditions have a substantial effect on several properties of a system, and this has been extensively studied in the Ising model. While they do not modify the bulk critical exponents, they lead to distinct boundary field theory and can change the finite-size scaling of the energy gap  and entanglement entropy~\cite{cabrera1987role, campostrini2015quantum, franchi2022critical, affleck2009entanglement}.

\begin{figure}[!htb]
    \centering
    \includegraphics[width=0.95\linewidth]{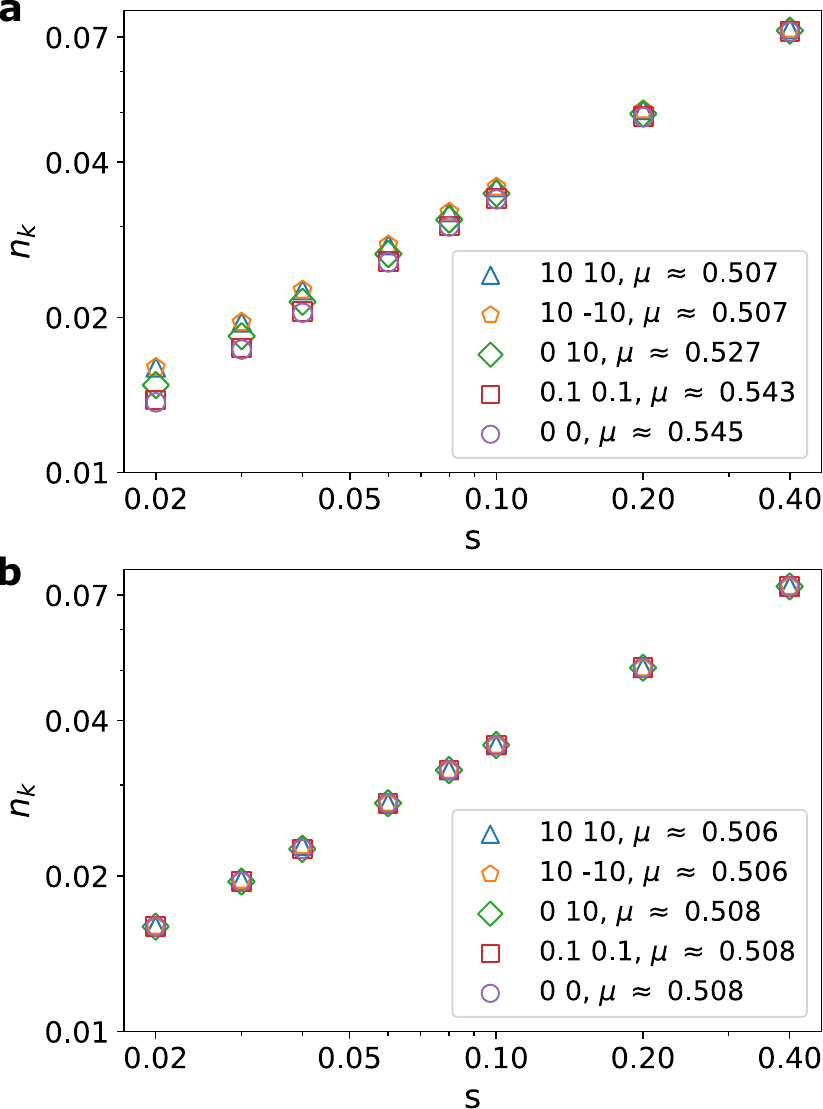}
    \caption{Scaling of the density of kinks $n_k$ with the sweep rate $s$ for the antiferromagnetic Ising model for various boundary conditions, for a system of $L = 201$ sites. The kink density is measured (a) over the entire chain and (b) over the central $10\%$ of the chain, using the whole set of data points shown in the figure. Boundary conditions are fixed with external longitudinal field applied at first and last site of the chain and indicated in the legends. The results evaluated in the central part of the chain always agree within 2$\%$ with the theory prediction $\mu=1/2$, while the results evaluated over the entire chain provide more accurate estimate of the KZ exponent when the boundary conditions are fixed.}
    \label{fig:IsingBoundary}
\end{figure}

\begin{figure}[!htb]
    \centering
    \includegraphics[width=0.95\linewidth]{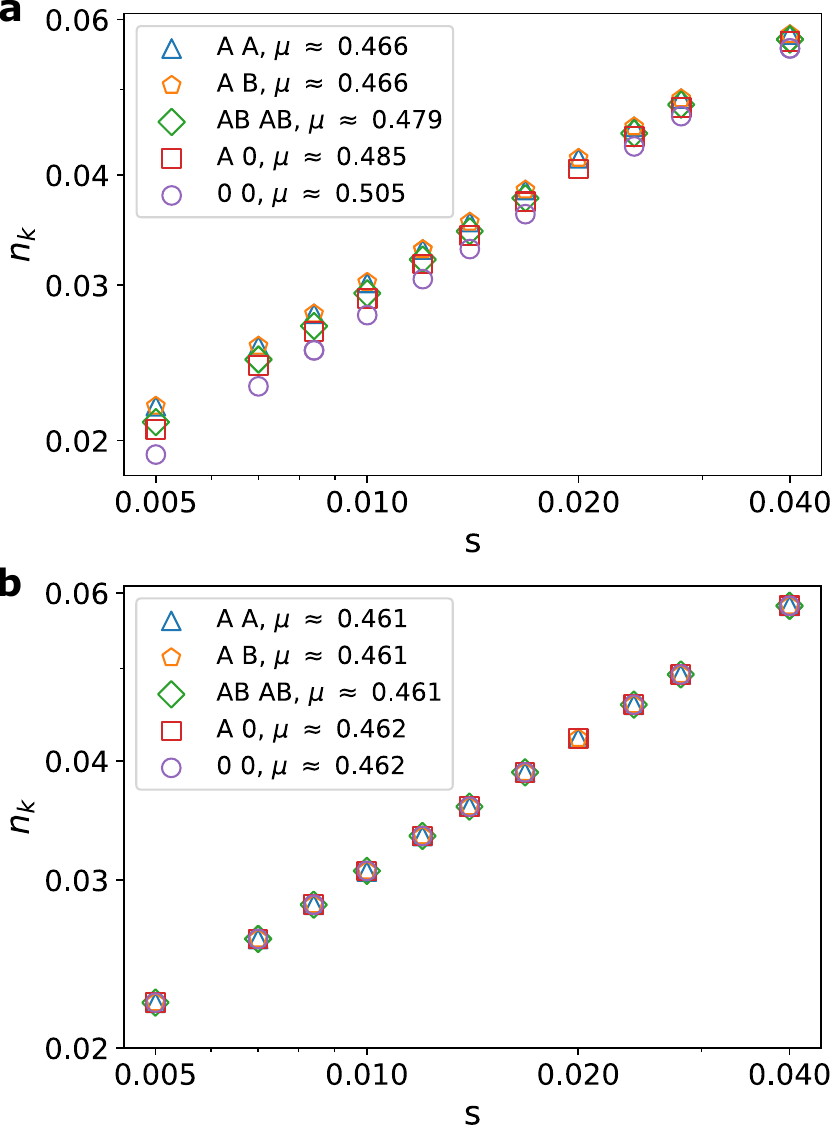}
    \caption{Scaling of the density of kinks $n_k$ with the sweep rate $s$, in the Potts model for various boundary conditions when the density of kinks is measured (a) over the whole chain, and (b) over the central $10\%$ interval of the chain for the whole set of data point depicted in the figure. Fixed boundary conditions are labeled with A or B, mixed with AB (not-C), free boundary is marked with 0. The results evaluated in the central part of the chain always agree within 2$\%$ with the theory prediction $\mu=5/11$, while the results evaluated over the entire chain provide more accurate estimate of the KZ exponent when the boundary conditions are fixed.}
    \label{fig:PottsBoundary}
\end{figure}

Recently, the role of periodic and antiperiodic boundary conditions in the context of the KZ mechanism has also been investigated~\cite{gomez2022role}, where it was found that Kibble–Zurek scaling remains essentially unchanged whether the boundary conditions are periodic or antiperiodic. Here, we build on these findings by conducting a more detailed study of various boundary conditions. To investigate whether boundary conditions influence the apparent Kibble–Zurek exponent, we examined the scaling of kink densities for both the Ising and Potts models under various boundary conditions. The corresponding results are shown in Fig.~\ref{fig:IsingBoundary} and~\ref{fig:PottsBoundary}. We evaluate the kink density in two ways: (i) over the entire chain, including the edges, and (ii) within a small interval centered in the middle of the chain. Unless stated otherwise, all results in this section were obtained for $L = 201$, and kinks are defined according to the standard convention (see previous section for details). The boundaries are polarized by applying a longitudinal magnetic field, effectively imposing fixed boundary conditions.

We assess how boundary conditions affect the estimation of the critical exponent $\mu$ in Fig.~\ref{fig:IsingBoundary}. We consider five representative cases for the Ising model: 
(i) strongly polarized symmetric boundaries ($h_{z1}=h_{zL}=10$); 
(ii) strongly polarized antisymmetric boundaries ($h_{z1}=-h_{zL}=10$); 
(iii) weakly polarized symmetric boundaries ($h_{z1}=h_{zL}=0.1$); 
(iv) free boundaries ($h_{z1}=h_{zL}=0$); 
and (v) one end free and the other polarized ($h_{z1}=0$, $h_{zL}=10$). 
Table~\ref{tab:mu-bc} summarizes the corresponding estimates of $\mu$, obtained both by 
counting kinks along the whole chain and by restricting the count to a small central interval 
so as to suppress boundary-induced effects.
\begin{table}[t]
\centering
\footnotesize
\setlength{\tabcolsep}{12pt}
\begin{tabular}{lcc}
\hline\hline
Boundary Conditions & $\mu$ (whole) & $\mu$ (center) \\
\hline
Strong symmetric      & $0.507 \pm 0.008$ & $0.506 \pm 0.014$ \\
Strong antisymmetric  & $0.507 \pm 0.008$ & $0.506 \pm 0.014$ \\
Weak symmetric        & $0.527 \pm 0.008$ & $0.508 \pm 0.008$ \\
Free                  & $0.543 \pm 0.018$ & $0.508 \pm 0.008$ \\
Fixed-Free    & $0.545 \pm 0.021$ & $0.508 \pm 0.008$ \\
\hline\hline
\end{tabular}
\caption{KZ exponent $\mu$, for several boundary conditions (BC) in the Ising model. Whole-chain values show strong boundary dependence, while central-interval values converge to a universal bulk exponent. The errors come from the fitting, as explained in the Methods section. The theoretical prediction for the Ising transition is $\mu = 0.5$.}
\label{tab:mu-bc}
\end{table}

When kinks are counted in the small central interval ($\approx10\%$ of the total length) of a long chain, the density of kinks evaluated for all five boundary conditions scales with a KZ critical exponent that agrees within $2\%$ with the theory expectation $\mu=0.5$. If instead, the density of kinks is evaluated over the entire chain (as it is often the case in experiments where available system sizes are limited) the critical exponent evaluated with the fixed boundary conditions turns out to be significantly more accurate. Surprisingly, in the regime where the Kibble–Zurek mechanism exhibits power-law scaling, we find no discernible difference in the KZ dynamics between fixed symmetric and fixed antisymmetric boundary conditions. This behavior contrasts with the ground-state properties, where the energy gap scales differently with system size~\cite{cabrera1987role}.

For the ferromagnetic three-state Potts model, we  examine five different boundary conditions, shown in Fig.~\ref{fig:PottsBoundary}:
(i) fixed boundary conditions favoring the same ferromagnetic state at both edges (A~A); 
(ii) fixed boundary conditions favoring different states at the two edges (A~B); 
(iii) free boundaries (0~0); 
(iv) fixed--free boundaries (A~0); 
and (v) mixed boundaries (AB~AB), implemented by applying a longitudinal field in the $C$ direction at both edges. 
The first four cases are direct generalizations of the boundary conditions used for the Ising model. 
The corresponding estimates of $\mu$, obtained by counting kinks either along the whole chain or only within a small central interval, 
are summarized in Table~\ref{tab:mu-potts}.

\begin{table}[t]
\centering
\footnotesize
\setlength{\tabcolsep}{12pt}
\begin{tabular}{lcc}
\hline\hline
Boundary Conditions & $\mu$ (whole) & $\mu$ (center) \\
\hline
Fixed same (A A)      & $0.466 \pm 0.004$ & $0.461 \pm 0.005$ \\
Fixed different (A B) & $0.466 \pm 0.004$ & $0.461 \pm 0.005$ \\
Free (0 0)            & $0.479 \pm 0.006$ & $0.461 \pm 0.006$ \\
Fixed--free (A 0)     & $0.485 \pm 0.009$ & $0.462 \pm 0.004$ \\
Mixed (AB AB)         & $0.505 \pm 0.019$ & $0.462 \pm 0.006$ \\
\hline\hline
\end{tabular}
\caption{KZ exponent $\mu$ for the ferromagnetic three-state Potts model under different boundary conditions (BC). Whole-chain estimates depend strongly on the BC, while central-interval values converge to a consistent bulk exponent. The error estimation comes from the fitting, as explained in the Methods section. The theoretical prediction for the 3-state Potts transition is $\mu \approx 0.454$. }
\label{tab:mu-potts}
\end{table}

Similar to the Ising model case, we see that when the KZ exponent is extracted from the total density of kinks, the results obtained with fixed boundary conditions are significantly more accurate than those obtained with other types of boundaries (see Fig.~\ref{fig:PottsBoundary}(a)). If, however, the density of kinks is evaluated only in the central interval of a long enough chain, the boundary conditions do not matter, just like we have observed for the Ising model in Fig.~\ref{fig:IsingBoundary}(b). 

\begin{figure}[!htb]
    \centering
    \includegraphics[width=0.95\linewidth]{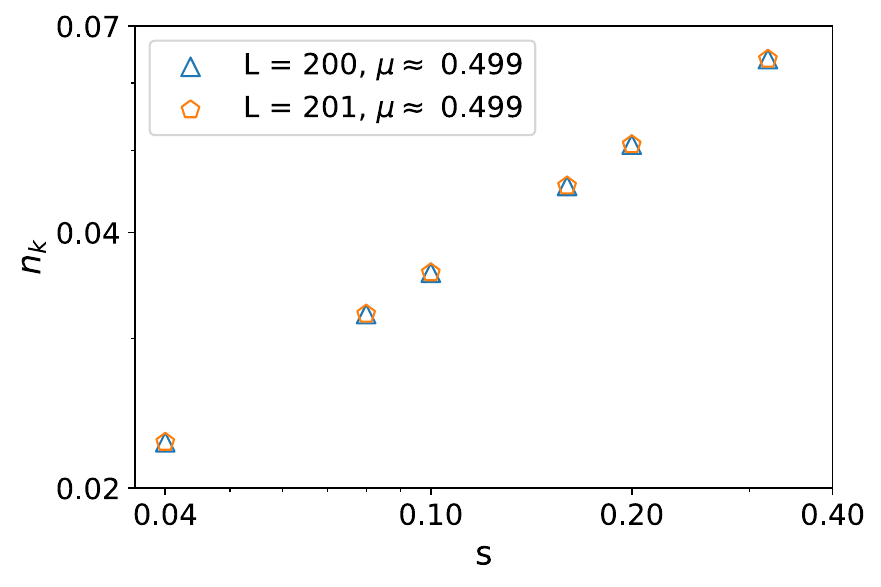}
    \caption{Comparison of the kink density scaling for the antiferromagnetic Ising model under periodic boundary conditions (PBC) when the system size  is compatible  (blue markers) and when it is incompatible (orange markers) with the periodicity of the ground state.The latter case is equivalent to imposing anti-periodic boundary conditions in the conformal field theory sense. The two data sets overlap almost perfectly, indicating that the scaling behavior is unaffected by the boundary choice.}
    \label{fig:TFIMpbc}
\end{figure}

We also examine the scenario of periodic (PBC) and anti-periodic boundary conditions (APBC) by considering antiferromagnetic Ising model on a closed loops with even and odd number of sites that are correspondingly compatible and incompatible with the ordered phase. In Fig.~\ref{fig:TFIMpbc}, we compare the scaling of the kink density for the antiferromagnetic Ising model under PBC for two system sizes, $L = 200$ (the system size compatible with the periodicity of the ground state) and $L = 201$ (the system size incompatible with the ground-state periodicity, often called anti-periodic boundary conditions in the field-theory context). In both cases, the extracted KZ critical exponent is $0.4989\pm0.0023$, indicating that $\mu$ is independent of this choice.

\begin{figure}[!htb]
    \centering
    \includegraphics[width=0.95\linewidth]{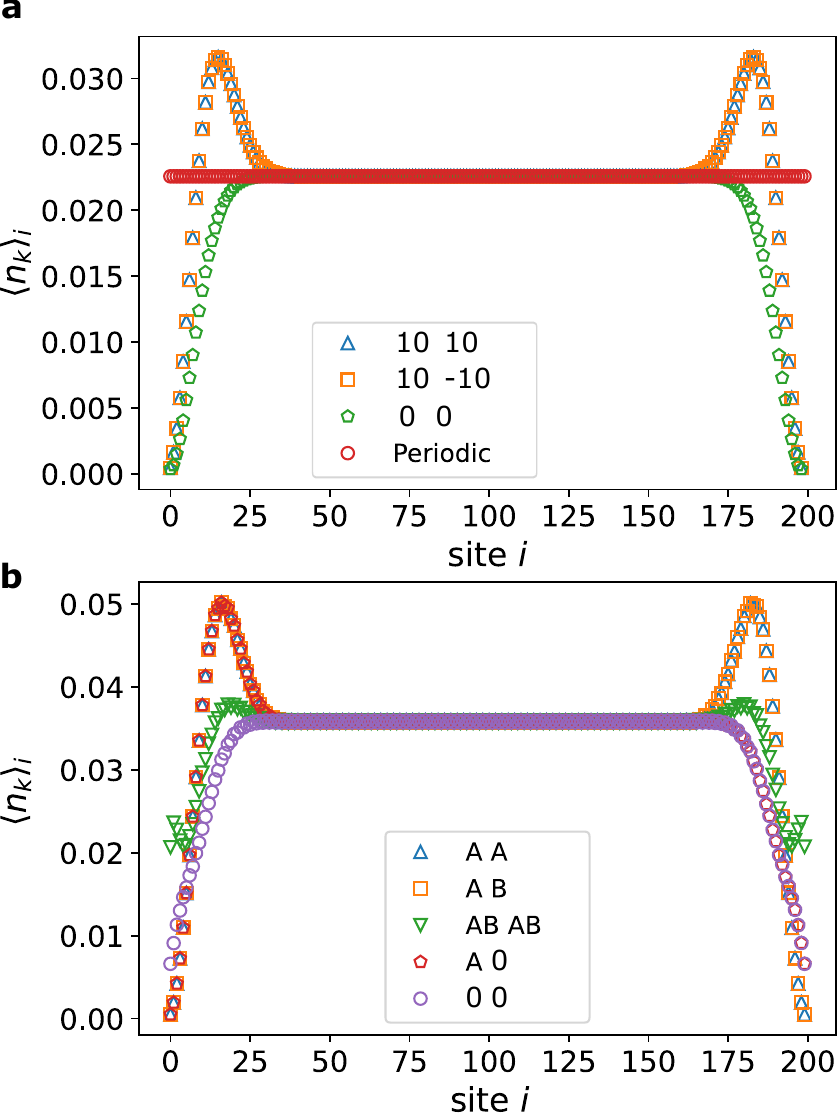}
    \caption{Distribution of kinks along the chain after a quench for (a) the Ising models at sweep rate $s = 0.04$ and (b) the Potts model at $s = 0.014$ for various boundary conditions listed in the legend.}
    \label{fig:kinksdistributionboundary}
\end{figure}

In Fig.~\ref{fig:kinksdistributionboundary}, we present the local kink density along the chain under various boundary conditions for the Ising model (Fig.~\ref{fig:kinksdistributionboundary}(a)) and the Potts model (Fig.~\ref{fig:kinksdistributionboundary}(b)). Both models exhibit remarkably similar kink distribution profiles when subjected to identical boundary conditions. For the periodic boundaries, the kink distribution, as expected, is uniform across the chain. For open and fixed boundaries---whether fixed in the same or opposite directions---kinks tend to accumulate near, but not directly adjacent to, the boundaries, reaching zero at the edges. Remarkably, although the total number of kinks is the same under fixed and periodic boundary conditions, their spatial distribution along the chain differs.

When boundaries are open but free, the peak in the kink distribution observed for fixed boundaries disappears, although the kinks density still decreases toward the boundaries. Under these conditions, the total number of kinks is reduced compared to the fixed boundary conditions. 

Finally, for the mixed boundaries in the Potts model---where the degree of freedom is partially constrained at the edges, such that local state C is disfavored, while A and B can appear with equal probability---the peak in the kink distribution is smaller than in the fixed case. Here, the total number of kinks falls between those for free and fixed boundary conditions. And, as we have seen in Fig. 
\ref{fig:PottsBoundary} the accuracy of the KZ exponent also lies in between the fixed and free cases. 

\begin{figure}[!htb]
    \centering
    \includegraphics[width=0.95\linewidth]{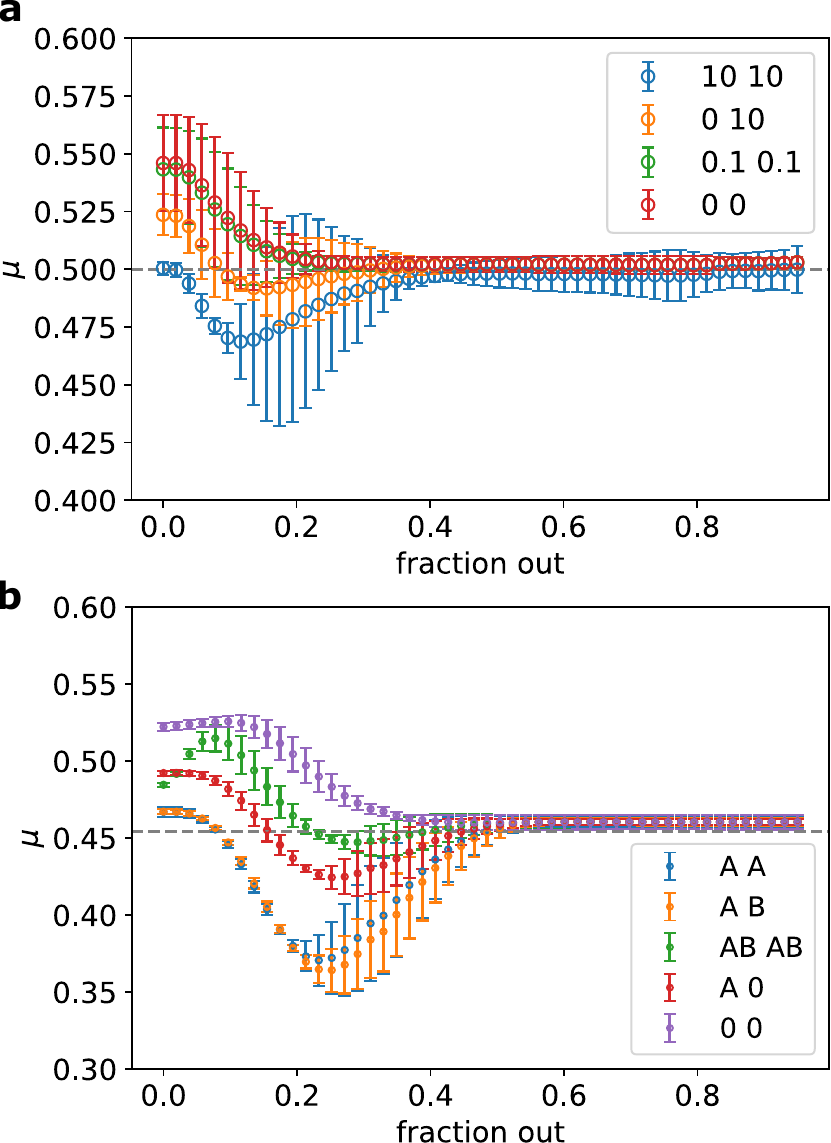}
    \caption{Kibble-Zurek critical exponent $\mu$ as a function of the fraction of the edge discarded for density calculations for various boundary conditions for (a) The Ising and (b) the three-state Potts model for a system size $L = 201$ sites. The exponent $\mu$ was calculated using the same range of sweep rates as in Fig.~\ref{fig:IsingBoundary} and Fig.~\ref{fig:PottsBoundary} respectively. Dashed lines indicate theory predictions $\mu=0.5$ for Ising and $\mu=5/11$ for Potts. Error bars represent the uncertainty in the exponent calculated as indicated in the methods section.}
    \label{fig:muvsfragment}
\end{figure}

The final question we address is how the extracted KZ exponent is affected by the size of the fragment retained in the calculation when different boundary conditions are used. Fig.~\ref{fig:muvsfragment} shows the value of the KZ exponent $\mu$ as a function of the fraction of sites discarded at each edge of a chain with $L = 201$ for several choices of boundary conditions. We selected the same range of $s$ as in the calculations shown in Fig.~\ref{fig:IsingBoundary} and Fig.~\ref{fig:PottsBoundary}. The kink density is computed only within the retained central interval, allowing us to quantify how sensitive the fitted exponent is to boundary effects.

Ising and Potts models show a similar behaviour. In both cases $\mu$ matches the theoretical value for the whole chain when boundaries are fixed. For the chosen system size $L = 201$ and for all boundary conditions, the edge effects seem to vanish after half of the considered chains are discarded. This interval is expected to be model and system size dependent, but already this simple examples of the two minimal models provides us with a good indication that one has to discard many dozens of sites posting quite a significant lower bound on the total length of the chain.

\begin{figure}[!htb]
    \centering
    \includegraphics[width=0.95\linewidth]{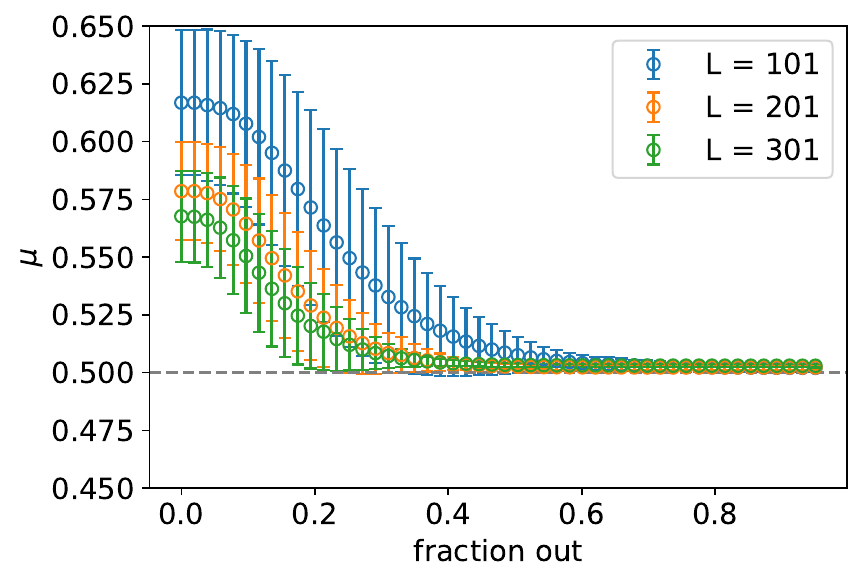}
    \caption{Kibble-Zurek critical exponent $\mu$ as a function of the fraction of sites discarded at each boundary when computing the kink density, shown for several system sizes of the Ising model. The exponent $\mu$ was extracted from the same range of sweep rates as in Fig.~\ref{fig:IsingBoundary}. Error bars indicate the uncertainty in the extracted exponent, evaluated as described in the Methods section.}
    \label{fig:fragmentsize}
\end{figure}

In Fig.~\ref{fig:fragmentsize} we perform the same comparison of the KZ exponent as a function of the size of the discarded fragment for the same range of $s$ as in Fig.~\ref{fig:IsingBoundary}, but for free boundary conditions $h_{z1} = h_{zN} = 0$ and various system sizes. Bigger system sizes yield a critical exponent $\mu$ closer to the theoretical $\mu = 0.5$ when the whole chain is considered and converge faster to the theoretical $\mu$ as a function of fraction discarded, although not faster as a function of the number of atoms discarded.

\section{Arrays of Rydberg atoms}
\label{sec:Rydberg}
Finally, we test whether kink definition also applies to arrays of Rydberg atoms. The Rydberg-atom system can be described by the Hamiltonian of interacting hard-core bosons
\begin{equation}
    H = \frac{\Omega}{2}\sum_i \left(d_i + d^\dagger_i\right)
    - \Delta \sum_i n_i
    + \sum_{i<j} V_{ij} n_i n_j,
    \label{eq:tail}
\end{equation}
where $\Omega$ is the Rabi frequency driving an atom to the Rydberg state, $\Delta$ is the laser detuning, and $V_{ij} \propto 1/r^6$ is the van der Waals interaction.

We perform a quench from the disordered phase to the period-2 phase at a constant rate. Throughout the quench we keep $\Omega = 1$ and the blockade radius $R_b = 1.5$ fixed, and vary the detuning from $\Delta = -5$ to $\Delta = 5$. Note that our quench protocol differs slightly from that used in Ref.~\cite{keesling2019quantum}. In their protocol, $\Omega$ is switched off before the measurement, making the ground state at the final point a perfect crystal. In our case $\Omega$ remains finite, and thus the ground state exhibits quantum fluctuations and is not perfectly ordered.

\begin{figure}
    \centering
    \includegraphics[width=\linewidth]{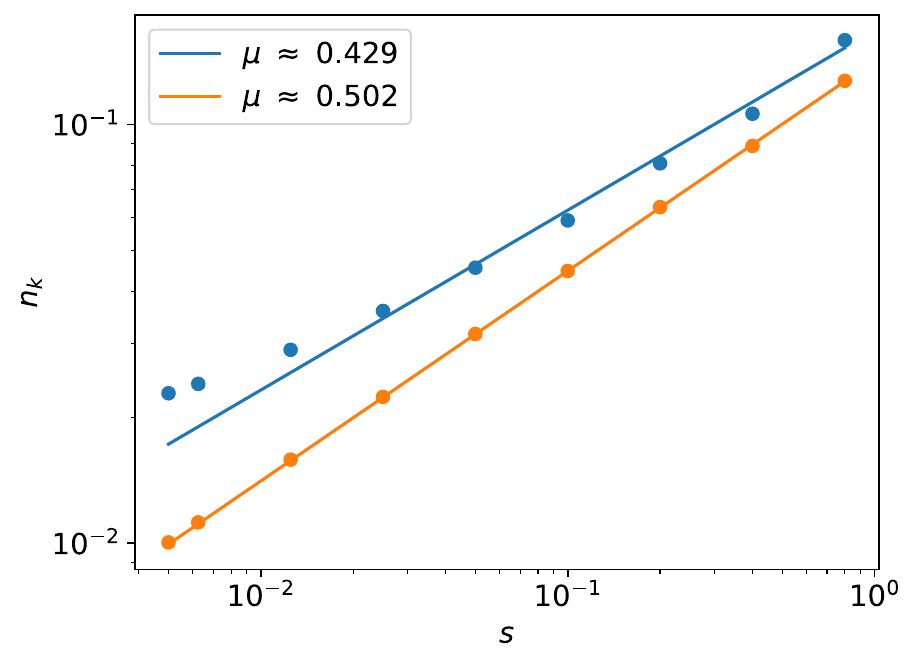}
    \caption{Scaling of the kink density $n_k$ with the sweep rate $s$ for the Rydberg model with van der Waals interactions decaying as $r^{-6}$. We quench to a final detuning $\Delta = 5$ while keeping $\Omega = 1$ and $R_b = 1.5$ fixed. Results are shown for the standard kink operator (blue) and for a kink operator that counts only isolated domain walls (orange), for a system size $L = 200$. All data points in the figure are included in the fit. The Kibble--Zurek exponent $\mu$ is extracted from the slope of the resulting power-law scaling.}

    \label{fig:rydberg}
\end{figure}

In Fig.~\ref{fig:rydberg} we compare two definitions of kinks: the standard period-2 kink 
$\emptycircle\emptycircle + \fullcircle\fullcircle$ (blue markers), and our proposed isolated-kink definition 
$\emptycircle\fullcircle\fullcircle\emptycircle + \fullcircle\emptycircle\emptycircle\fullcircle$ (orange markers). We consider a system of size $L = 200$, whose length does not match the periodicity of the underlying crystal. The extracted KZ scaling exponents are $\mu = 0.43 \pm 0.04$ for the standard kink definition and $\mu = 0.502 \pm 0.004$ for the isolated-kink definition, the latter agreeing with the theoretical Ising prediction within $0.4\%$.

These results demonstrate that, within the range of quench rates exhibiting clean KZ power-law scaling, it is not necessary to choose a system size that matches the crystal periodicity, and that isolated kinks provide a more robust measure in the presence of quantum fluctuations of the ground state.

\section{Conclusion and Discussion}
\label{sec:Conclusion}
In this paper, we have studied the effect of boundary conditions and ending point in the Kibble-Zurek mechanism.

Our main finding is that the accuracy of the Kibble-Zurek mechanism in the naive implementation is extremely sensitive to the location of the final point---the point where the time evolution terminates and the measurements are performed. Our data show that, for the two minimal models considered here, the best choice of the final point would be the point where the transverse field responsible for quantum fluctuations vanishes.  If the final measurement point deviates from this classical limit, whether it lies before or after the zero-transverse-field point, the observed critical exponent $\mu$ deviates markedly from the theoretically predicted universal value.

Interestingly enough, exactly the same type of fine-tuning to the classical end point has been realized in Rydberg simulators. In the experiment, after quenching through a quantum phase transition, the Rabi frequency $\Omega$ responsible for quantum fluctuations of atoms between the ground-state and excited Rydberg state has to be turned off to perform the read-outs. The only difference between the two protocols is a path to reach this specific end point. In the minimal models considered above, we quenched the transverse field itself to the point when it vanishes. In experiments, one quenches either the laser detuning (this would corresponds to an additional longitudinal field in the minimal models) or the combination of the detuning and Rabi frequency to an arbitrary point inside the ordered phase and then turns off the lasers responsible for the resonance. Importantly, in both protocols, the transition is crossed with a linear ramp. Numerical simulations that closely reproduce the experimental protocol demonstrate excellent agreement with the expected universal scaling at the Ising and Potts points\cite{soto2024resolving}.

Such fine-tuning of the final point is natural for experiments on Rydberg atoms, but may not be generic in other contexts. In this paper, we present an alternative approach based on a more careful definition of the kink operator, which demonstrates a remarkable robustness to  the  location of the final endpoint. In this approach, the kink represents a domain wall between two {\it extended} domains, excluding, in particular, localized excitations within the domain of the same ground-state, like a single-spin flip. This definition requires some level of book-keeping but admits a conceptually straightforward generalization to other more complex models. 

In addition, we investigated how different boundary conditions affect the accuracy of the critical scaling. For both minimal models considered, we find that the accuracy of the Kibble-Zurek critical exponent is systematically improved for fixed boundary conditions with respect to the free ones. Surprisingly, we observe no difference in critical dynamics performed with symmetric and anti-symmetric (in the conformal field theory sense) fixed boundary conditions, despite their drastic differences in equilibrium scenarios. In effective models of Rydberg atoms, the presence of a longitudinal field (arising from laser detuning) partially fixes the boundary conditions by favoring the excitation of edge atoms into Rydberg states.  Increasing the laser detuning at the edges can further reinforce these effective fixed boundary conditions. The lack of any difference between symmetric and anti-symmetric boundary conditions relaxes experimental constraints. Typically, arrays with a total size $N=kp+1$ with $k\in\mathbb{Z}$ have been used to probe the transition into the period-$p$ phase. Our results show that the scaling should be universal for any values of $N$ (of course, $N$ has to be big enough to host the Kibble-Zurek regime).

We also report that, for sufficiently long chains, the universal Kibble-Zurek scaling can be accurately extracted by calculating the density of kinks in the central part of the chain, thus ignoring the effect of the boundaries. The size of the chain in this case must be substantially larger than the resulting correlation length after a quench, which naturally increases with decreasing sweep rate. For the minimal models and sweep rates considered here, the edge effects are significant over the first 30-40 sites, imposing a lower bound on the total chain length of at least $N\gtrsim100$ sites. These lengths are feasible but challenging in modern Rydberg experiments, so fixing the boundary conditions seems to be the best strategy for now.

\section*{Acknowledgements}
We acknowledge useful discussions with Rhine Samajdar and Hannes Bernien.
  This research has been supported by Delft Technology Fellowship (NC).  
  Numerical simulations have been performed at the DelftBlue HPC and at the Dutch national e-infrastructure with the support of the SURF Cooperative.
\begin{appendix}
\section{Convergence and bond dimension}
\begin{figure}[tb]
    \centering
    \includegraphics[width=0.95\linewidth]{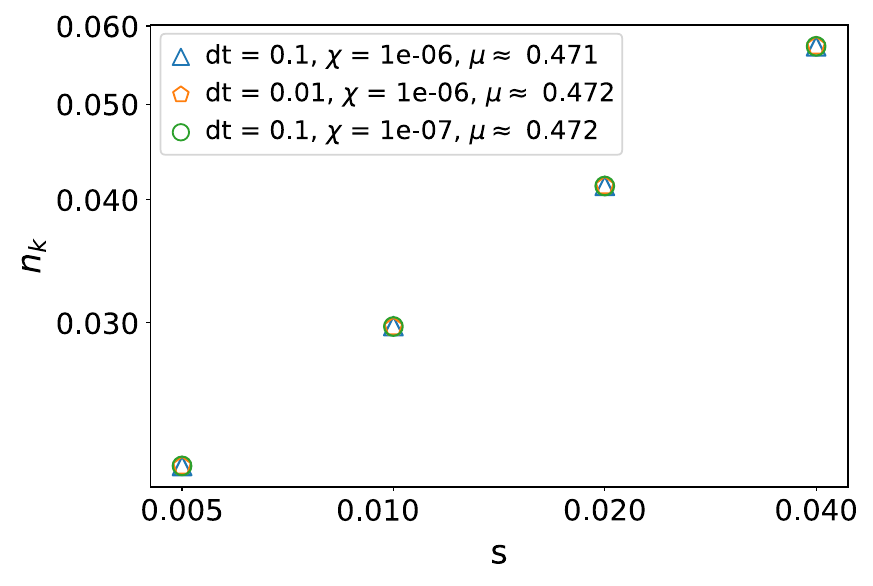}
    \caption{Comparison of kink density for different numerical parameters in the Potts model with system size $L = 101$. We test two values of the singular value cutoff ($\chi = 10^{-6}$ and $\chi = 10^{-7}$) and two time step values ($\textnormal{dt} = 0.1$ and $\textnormal{dt} = 0.01$). The results show negligible differences across these parameters, indicating the robustness of the kink density against moderate variations in truncation threshold and time discretization.}
    \label{fig:pottsparameters}
\end{figure}

In Fig.~\ref{fig:pottsparameters}, we compare two cutoff values for the minimum singular value, $\chi = 10^{-6}$ and $\chi = 10^{-7}$, as well as two time step values, $\textnormal{dt} = 0.1$ and $\textnormal{dt} = 0.01$. We observe no significant differences in the kink density across these parameters.

\begin{figure}[!h]
    \centering
    \includegraphics[width=0.95\linewidth]{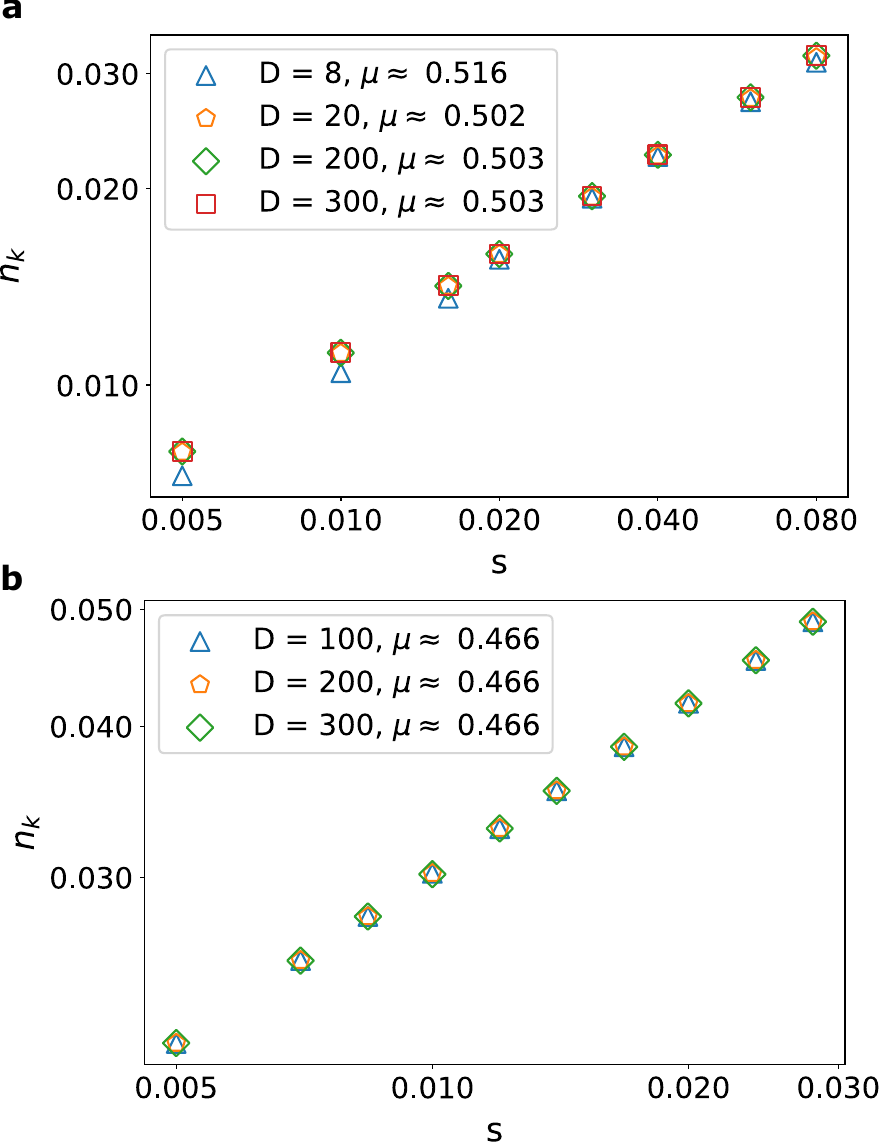}
    \caption{Kibble-Zurek scaling of the density of kinks $n_k$ as a function of the sweep rate for various bond dimensions $D$ in (a) the 1D transverse-field Ising model and (b) the 1D quantum 3-state Potts model.In both cases, $ L = 201$ sites. Simulation of Kibble-Zurek dynamics requires relatively small bond dimension ($D\lesssim 100$). For the rest of the simulations $D = 300$ was used for both Ising Potts models.}
    \label{fig:D}
\end{figure}

Fig.~\ref{fig:D} presents a comparison of various bond dimensions $D$ for both the Ising model (Fig.~\ref{fig:D}(a)) and the Potts model (Fig.~\ref{fig:D}(b)). In both cases, the Kibble-Zurek exponent $\mu$ remains stable for moderately large $D$, with noticeable deviations occurring only at very small bond dimensions. Based on these observations, we fixed the time step at $\delta = 0.1$, set the maximum bond dimension to $D = 300$, and applied a singular value cutoff of $\chi > 10^{-6}$ for the simulations presented in the main text.

For simulations employing periodic boundary conditions, the time-dependent variational principle (TDVP) \cite{haegeman2011time, haegeman2016unifying, paeckel2019time} was used instead, while retaining the same values of $D$ and $\chi$ as in TEBD.

\section{Kink density profiles}
\begin{figure}[htb]
    \centering
    \includegraphics[width=0.9\linewidth]{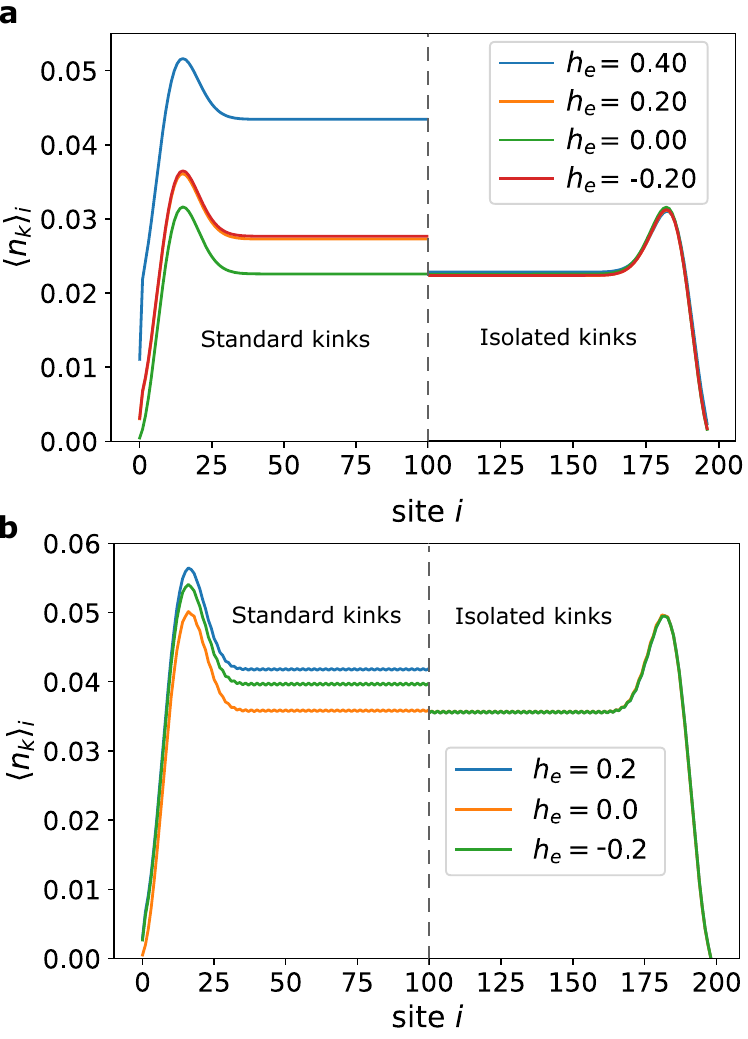}
    \caption{Local density of kinks $n_k$ along the chain for (a) the Ising and (b) the Potts models for various values of the final transverse field $h_e$ and finite system size $L=201$. The sweep rates values considered in the figure are $s = 0.04$ for Ising model and $s = 0.014$ for Potts. Each plot is split in two: in the left side, the kink operator is the standard domain wall operator, while in the right side only isolated kinks are measured. For both, Ising and Potts models, when $h_e=0$ the profiles of standard and isolated kinks coincide. However, for different end points the profile of standard kinks is shifted, while the profile for isolated kinks remains robust.}
    \label{fig:kinksdistribution}
\end{figure}

In Fig.~\ref{fig:kinksdistribution}, we show the profiles of the local kink density value for the Ising and models at various endpoints of a single sweep rate and finite system size. The operator local kink operator is just the density of kinks operator applied to a single location of the chain. A direct comparison between standard and isolated kinks reveals that all profiles corresponding to the same sweep rate and system size, but ending at different final fields, are nearly identical up to a constant vertical shift. This shift appears to result from single-spin-flip excitations, as it disappears entirely when considering isolated kinks: in that case, all curves collapse onto each other.

\begin{figure}[!htb]
    \centering
    \includegraphics[width=0.9\linewidth]{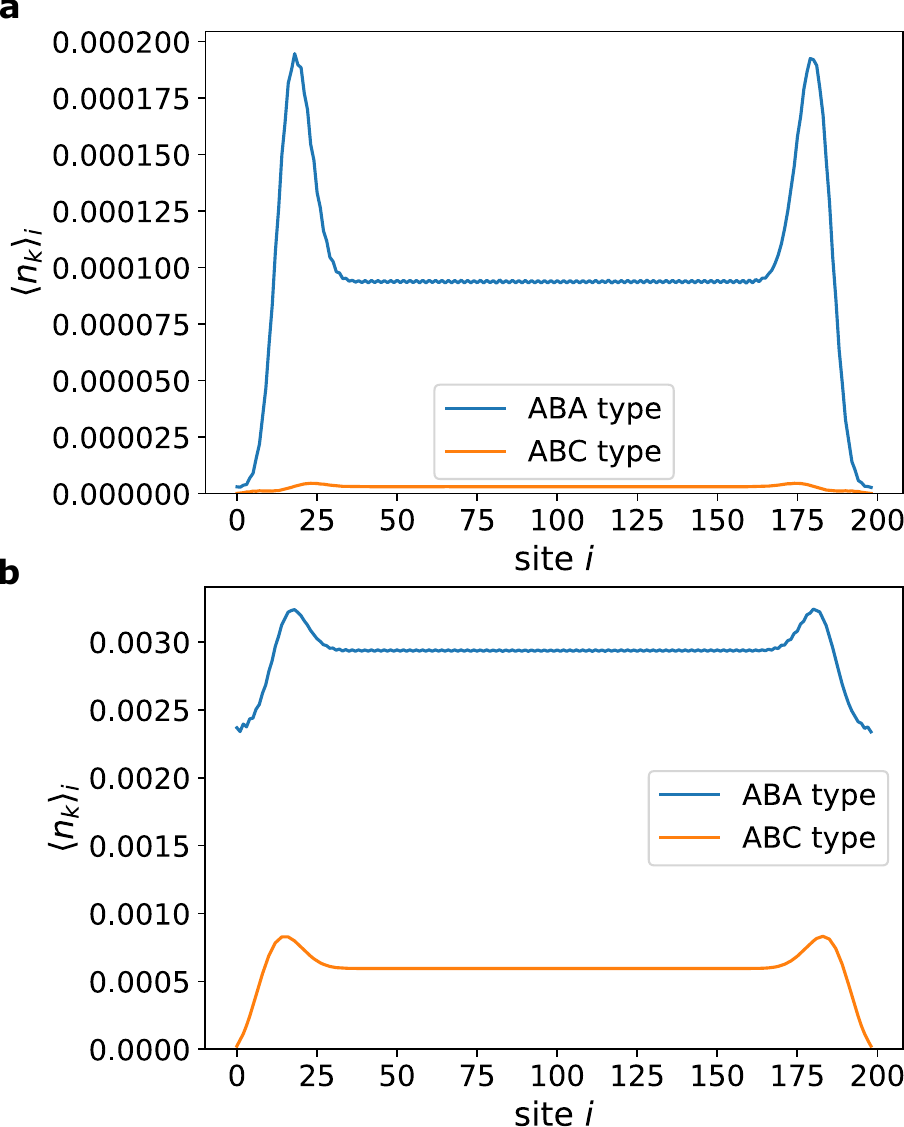}
    \caption{Local density of two types of local kinks $n_k$ in the Potts model after a slow quench with sweep rate $s=0.014$, shown along a finite chain with $L = 201$ sites for (a) $h_e=0$ and (b) $h_e=0.2$. The blue curve corresponds to spin flips inside a domain, and the orange curve to spin flips at a domain boundary. In both cases spin flips inside a domain are more abundant than at a domain boundary.}
    \label{fig:alternativekink}
\end{figure}

In the Potts model, we can also compare the local density of spin flips inside or between domains for $h_e = 0$ and $h_e \neq 0$. Fig.~\ref{fig:alternativekink} illustrates these scenarios: panel (a) shows the local density profiles for both types of spin flips when $h_e = 0$, and panel (b) shows them when $h_e = 0.2$. As expected from Fig.~\ref{fig:kinksdistribution}, spin flips occur more frequently under nonzero $h_e$. Moreover, even though flipping a spin inside a domain costs twice as much energy as flipping one at a domain boundary, such “in-domain” flips are still more common. The apparent contradiction is resolved by noting that domain walls are relatively rare, making it statistically more likely for a random spin to be inside a domain rather than at its boundary.

\end{appendix}
\clearpage
\bibliographystyle{unsrt}
\bibliography{bibliography}

\begin{thebibliography}{10}

\bibitem{sachdev2011quantum}
Subir Sachdev.
\newblock {\em Quantum Phase Transitions}.
\newblock Cambridge University Press, 2 edition, 2011.

\bibitem{cardy1996scaling}
John Cardy.
\newblock {\em Scaling and renormalization in statistical physics}, volume~5.
\newblock Cambridge university press, 1996.

\bibitem{zurek2005dynamics}
Wojciech~H Zurek, Uwe Dorner, and Peter Zoller.
\newblock Dynamics of a quantum phase transition.
\newblock {\em Physical review letters}, 95(10):105701, 2005.

\bibitem{polkovnikov2005}
Anatoli Polkovnikov.
\newblock Universal adiabatic dynamics in the vicinity of a quantum critical point.
\newblock {\em Phys. Rev. B}, 72:161201, Oct 2005.

\bibitem{uhlmann2007vortex}
Michael Uhlmann, Ralf Sch{\"u}tzhold, and Uwe~R Fischer.
\newblock Vortex quantum creation and winding number scaling in a quenched spinor bose gas.
\newblock {\em Physical review letters}, 99(12):120407, 2007.

\bibitem{dziarmaga2010dynamics}
Jacek Dziarmaga.
\newblock Dynamics of a quantum phase transition and relaxation to a steady state.
\newblock {\em Advances in Physics}, 59(6):1063--1189, 2010.

\bibitem{keesling2019quantum}
Alexander Keesling, Ahmed Omran, Harry Levine, Hannes Bernien, Hannes Pichler, Soonwon Choi, Rhine Samajdar, Sylvain Schwartz, Pietro Silvi, Subir Sachdev, et~al.
\newblock Quantum kibble--zurek mechanism and critical dynamics on a programmable rydberg simulator.
\newblock {\em Nature}, 568(7751):207--211, 2019.

\bibitem{uhlmann2010n}
Michael Uhlmann, Ralf Sch{\"u}tzhold, and Uwe~R Fischer.
\newblock O (n) symmetry-breaking quantum quench: Topological defects versus quasiparticles.
\newblock {\em Physical Review D—Particles, Fields, Gravitation, and Cosmology}, 81(2):025017, 2010.

\bibitem{uhlmann2010system}
Michael Uhlmann, Ralf Sch{\"u}tzhold, and Uwe~R Fischer.
\newblock System size scaling of topological defect creation in a second-order dynamical quantum phase transition.
\newblock {\em New Journal of Physics}, 12(9):095020, 2010.

\bibitem{del2014universality}
Adolfo Del~Campo and Wojciech~H Zurek.
\newblock Universality of phase transition dynamics: Topological defects from symmetry breaking.
\newblock {\em International Journal of Modern Physics A}, 29(08):1430018, 2014.

\bibitem{kibble1976topology}
Thomas~WB Kibble.
\newblock Topology of cosmic domains and strings.
\newblock {\em Journal of Physics A: Mathematical and General}, 9(8):1387, 1976.

\bibitem{kibble1980some}
Tom~WB Kibble.
\newblock Some implications of a cosmological phase transition.
\newblock {\em Physics Reports}, 67(1):183--199, 1980.

\bibitem{zurek1985cosmological}
Wojciech~H Zurek.
\newblock Cosmological experiments in superfluid helium?
\newblock {\em Nature}, 317(6037):505--508, 1985.

\bibitem{dziarmaga2005dynamics}
Jacek Dziarmaga.
\newblock Dynamics of a quantum phase transition: Exact solution of the quantum ising model.
\newblock {\em Physical review letters}, 95(24):245701, 2005.

\bibitem{kolodrubetz2012nonequilibrium}
Michael Kolodrubetz, Bryan~K Clark, and David~A Huse.
\newblock Nonequilibrium dynamic critical scaling of the quantum ising chain.
\newblock {\em Physical review letters}, 109(1):015701, 2012.

\bibitem{chandran2012kibble}
Anushya Chandran, Amir Erez, Steven~S Gubser, and Shivaji~L Sondhi.
\newblock Kibble-zurek problem: Universality and the scaling limit.
\newblock {\em Physical Review B—Condensed Matter and Materials Physics}, 86(6):064304, 2012.

\bibitem{francuz2016space}
Anna Francuz, Jacek Dziarmaga, Bart{\l}omiej Gardas, and Wojciech~H Zurek.
\newblock Space and time renormalization in phase transition dynamics.
\newblock {\em Physical Review B}, 93(7):075134, 2016.

\bibitem{sadhukhan2020sonic}
Debasis Sadhukhan, Aritra Sinha, Anna Francuz, Justyna Stefaniak, Marek~M Rams, Jacek Dziarmaga, and Wojciech~H Zurek.
\newblock Sonic horizons and causality in phase transition dynamics.
\newblock {\em Physical Review B}, 101(14):144429, 2020.

\bibitem{bernien2017probing}
Hannes Bernien, Sylvain Schwartz, Alexander Keesling, Harry Levine, Ahmed Omran, Hannes Pichler, Soonwon Choi, Alexander~S Zibrov, Manuel Endres, Markus Greiner, et~al.
\newblock Probing many-body dynamics on a 51-atom quantum simulator.
\newblock {\em Nature}, 551(7682):579--584, 2017.

\bibitem{laguna1997density}
Pablo Laguna and Wojciech~Hubert Zurek.
\newblock Density of kinks after a quench: When symmetry breaks, how big are the pieces?
\newblock {\em Physical Review Letters}, 78(13):2519, 1997.

\bibitem{saito2007kibble}
Hiroki Saito, Yuki Kawaguchi, and Masahito Ueda.
\newblock Kibble-zurek mechanism in a quenched ferromagnetic bose-einstein condensate.
\newblock {\em Physical Review A—Atomic, Molecular, and Optical Physics}, 76(4):043613, 2007.

\bibitem{de2010spontaneous}
Gabriele De~Chiara, Adolfo Del~Campo, Giovanna Morigi, Martin~B Plenio, and Alex Retzker.
\newblock Spontaneous nucleation of structural defects in inhomogeneous ion chains.
\newblock {\em New Journal of Physics}, 12(11):115003, 2010.

\bibitem{del2010structural}
A~Del~Campo, Gabriele De~Chiara, Giovanna Morigi, Martin~B Plenio, and A~Retzker.
\newblock Structural defects in ion chains by quenching the external potential:<? format?> the inhomogeneous kibble-zurek mechanism.
\newblock {\em Physical review letters}, 105(7):075701, 2010.

\bibitem{jaschke2017critical}
Daniel Jaschke, Kenji Maeda, Joseph~D Whalen, Michael~L Wall, and Lincoln~D Carr.
\newblock Critical phenomena and kibble--zurek scaling in the long-range quantum ising chain.
\newblock {\em New Journal of Physics}, 19(3):033032, 2017.

\bibitem{soto2024resolving}
Jose Soto~Garcia and Natalia Chepiga.
\newblock Resolving chiral transitions in one-dimensional rydberg arrays with quantum kibble-zurek mechanism and finite-time scaling.
\newblock {\em Physical Review B}, 110(12):125113, 2024.

\bibitem{weiler2008spontaneous}
Chad~N Weiler, Tyler~W Neely, David~R Scherer, Ashton~S Bradley, Matthew~J Davis, and Brian~P Anderson.
\newblock Spontaneous vortices in the formation of bose--einstein condensates.
\newblock {\em Nature}, 455(7215):948--951, 2008.

\bibitem{griffin2012multiferroics}
Sin{\'e}ad~M Griffin, Martin Lilienblum, Kris~T Delaney, Yu~Kumagai, Manfred Fiebig, and Nicola~A Spaldin.
\newblock Scaling behavior and beyond equilibrium in the hexagonal manganites.
\newblock {\em Physical Review X}, 2(4):041022, 2012.

\bibitem{lamporesi2013spontaneous}
Giacomo Lamporesi, Simone Donadello, Simone Serafini, Franco Dalfovo, and Gabriele Ferrari.
\newblock Spontaneous creation of kibble--zurek solitons in a bose--einstein condensate.
\newblock {\em Nature Physics}, 9(10):656--660, 2013.

\bibitem{mielenz2013trapping}
Manuel Mielenz, J~Brox, Steffen Kahra, G{\"u}nther Leschhorn, Magnus Albert, Tobias Sch{\"a}tz, Haggai Landa, and Benni Reznik.
\newblock Trapping of topological-structural defects in coulomb crystals.
\newblock {\em Physical review letters}, 110(13):133004, 2013.

\bibitem{pyka2013topological}
K~Pyka, J~Keller, HL~Partner, R~Nigmatullin, T~Burgermeister, DM~Meier, K~Kuhlmann, A~Retzker, Martin~B Plenio, WH~Zurek, et~al.
\newblock Topological defect formation and spontaneous symmetry breaking in ion coulomb crystals.
\newblock {\em Nature communications}, 4(1):2291, 2013.

\bibitem{ulm2013observation}
S~Ulm, J~Ro{\ss}nagel, G~Jacob, C~Deg{\"u}nther, ST~Dawkins, UG~Poschinger, R~Nigmatullin, A~Retzker, MB~Plenio, F~Schmidt-Kaler, et~al.
\newblock Observation of the kibble--zurek scaling law for defect formation in ion crystals.
\newblock {\em Nature communications}, 4(1):2290, 2013.

\bibitem{beugnon2017exploring}
J{\'e}r{\^o}me Beugnon and Nir Navon.
\newblock Exploring the kibble--zurek mechanism with homogeneous bose gases.
\newblock {\em Journal of Physics B: Atomic, Molecular and Optical Physics}, 50(2):022002, 2017.

\bibitem{ko2019kibble}
Bumsuk Ko, Jee~Woo Park, and Y-i Shin.
\newblock Kibble--zurek universality in a strongly interacting fermi superfluid.
\newblock {\em Nature physics}, 15(12):1227--1231, 2019.

\bibitem{cui2020experimentally}
Jin-Ming Cui, Fernando~Javier G{\'o}mez-Ruiz, Yun-Feng Huang, Chuan-Feng Li, Guang-Can Guo, and Adolfo del Campo.
\newblock Experimentally testing quantum critical dynamics beyond the kibble--zurek mechanism.
\newblock {\em Communications Physics}, 3(1):44, 2020.

\bibitem{bando2020probing}
Yuki Bando, Yuki Susa, Hiroki Oshiyama, Naokazu Shibata, Masayuki Ohzeki, Fernando~Javier G{\'o}mez-Ruiz, Daniel~A Lidar, Sei Suzuki, Adolfo Del~Campo, and Hidetoshi Nishimori.
\newblock Probing the universality of topological defect formation in a quantum annealer: Kibble-zurek mechanism and beyond.
\newblock {\em Physical Review Research}, 2(3):033369, 2020.

\bibitem{lee2024universal}
Kyuhwan Lee, Sol Kim, Taehoon Kim, and Y~Shin.
\newblock Universal kibble--zurek scaling in an atomic fermi superfluid.
\newblock {\em Nature Physics}, 20(10):1570--1574, 2024.

\bibitem{rigol2008thermalization}
Marcos Rigol, Vanja Dunjko, and Maxim Olshanii.
\newblock Thermalization and its mechanism for generic isolated quantum systems.
\newblock {\em Nature}, 452(7189):854--858, 2008.

\bibitem{del2022locality}
Adolfo del Campo, Fernando~Javier G{\'o}mez-Ruiz, and Hai-Qing Zhang.
\newblock Locality of spontaneous symmetry breaking and universal spacing distribution of topological defects formed across a phase transition.
\newblock {\em Physical Review B}, 106(14):L140101, 2022.

\bibitem{samajdar2024quantum}
Rhine Samajdar and David~A Huse.
\newblock Quantum and classical coarsening and their interplay with the kibble-zurek mechanism.
\newblock {\em arXiv preprint arXiv:2401.15144}, 2024.

\bibitem{difrancesco}
P.~Di~Francesco, P.~Mathieu, and D.~S{\'e}n{\'e}chal.
\newblock {\em Conformal Field Theory}.
\newblock Graduate Texts in Contemporary Physics. Springer, New York, 1997.

\bibitem{white1992density}
Steven~R White.
\newblock Density matrix formulation for quantum renormalization groups.
\newblock {\em Physical review letters}, 69(19):2863, 1992.

\bibitem{schollwock2011density}
Ulrich Schollw{\"o}ck.
\newblock The density-matrix renormalization group in the age of matrix product states.
\newblock {\em Annals of physics}, 326(1):96--192, 2011.

\bibitem{vidal2004efficient}
Guifr{\'e} Vidal.
\newblock Efficient simulation of one-dimensional quantum many-body systems.
\newblock {\em Physical review letters}, 93(4):040502, 2004.

\bibitem{verstraete2004matrix}
Frank Verstraete, Juan~J Garcia-Ripoll, and Juan~Ignacio Cirac.
\newblock Matrix product density operators: Simulation of finite-temperature and dissipative systems.
\newblock {\em Physical review letters}, 93(20):207204, 2004.

\bibitem{paeckel2019time}
Sebastian Paeckel, Thomas K{\"o}hler, Andreas Swoboda, Salvatore~R Manmana, Ulrich Schollw{\"o}ck, and Claudius Hubig.
\newblock Time-evolution methods for matrix-product states.
\newblock {\em Annals of Physics}, 411:167998, 2019.

\bibitem{haegeman2011time}
Jutho Haegeman, J~Ignacio Cirac, Tobias~J Osborne, Iztok Pi{\v{z}}orn, Henri Verschelde, and Frank Verstraete.
\newblock Time-dependent variational principle for quantum lattices.
\newblock {\em Physical review letters}, 107(7):070601, 2011.

\bibitem{haegeman2016unifying}
Jutho Haegeman, Christian Lubich, Ivan Oseledets, Bart Vandereycken, and Frank Verstraete.
\newblock Unifying time evolution and optimization with matrix product states.
\newblock {\em Physical Review B}, 94(16):165116, 2016.

\bibitem{pirvu2010matrix}
Bogdan Pirvu, Valentin Murg, J~Ignacio Cirac, and Frank Verstraete.
\newblock Matrix product operator representations.
\newblock {\em New Journal of Physics}, 12(2):025012, 2010.

\bibitem{xia2021kibble}
Chuan-Yin Xia and Hua-Bi Zeng.
\newblock Kibble zurek mechanism in rapidly quenched phase transition dynamics.
\newblock {\em arXiv preprint arXiv:2110.07969}, 2021.

\bibitem{zeng2023universal}
Hua-Bi Zeng, Chuan-Yin Xia, and Adolfo Del~Campo.
\newblock Universal breakdown of kibble-zurek scaling in fast quenches across a phase transition.
\newblock {\em Physical Review Letters}, 130(6):060402, 2023.

\bibitem{Note1}
At the boundaries between the KZ regime and the other two regimes, we observe a transient behavior in which the kinks density increases more rapidly with the sweep rate than the KZ scaling would predict. The transient regime between the fast-quench and the KZ regime is the pre-saturated regime, which occurs when the initial point is in the adiabatic regime, but the ending point is inside the non-adiabatic regime \cite {kou2023varying}. The size of both transient regimes is model dependent.

\bibitem{Dziarmaga2022kink}
Jacek Dziarmaga and Marek~M Rams.
\newblock Kink correlations, domain-size distribution, and emptiness formation probability after a kibble-zurek quench in the quantum ising chain.
\newblock {\em Physical Review B}, 106(1):014309, 2022.

\bibitem{sen2010quenching}
Diptiman Sen and Smitha Vishveshwara.
\newblock Quenching across quantum critical points: Role of topological patterns.
\newblock {\em Europhysics Letters}, 91(6):66009, 2010.

\bibitem{sharma2015one}
Shraddha Sharma and Amit Dutta.
\newblock One-and two-dimensional quantum models: Quenches and the scaling of irreversible entropy.
\newblock {\em Physical Review E}, 92(2):022108, 2015.

\bibitem{hegde2015quench}
Suraj Hegde, Vasudha Shivamoggi, Smitha Vishveshwara, and Diptiman Sen.
\newblock Quench dynamics and parity blocking in majorana wires.
\newblock {\em New Journal of Physics}, 17(5):053036, 2015.

\bibitem{cabrera1987role}
GG~Cabrera and R~Jullien.
\newblock Role of boundary conditions in the finite-size ising model.
\newblock {\em Physical Review B}, 35(13):7062, 1987.

\bibitem{campostrini2015quantum}
Massimo Campostrini, Andrea Pelissetto, and Ettore Vicari.
\newblock Quantum ising chains with boundary fields.
\newblock {\em Journal of Statistical Mechanics: Theory and Experiment}, 2015(11):P11015, 2015.

\bibitem{franchi2022critical}
Alessio Franchi, Davide Rossini, and Ettore Vicari.
\newblock Critical crossover phenomena driven by symmetry-breaking defects at quantum transitions.
\newblock {\em Physical Review E}, 105(3):034139, 2022.

\bibitem{affleck2009entanglement}
Ian Affleck, Nicolas Laflorencie, and Erik~S S{\o}rensen.
\newblock Entanglement entropy in quantum impurity systems and systems with boundaries.
\newblock {\em Journal of Physics A: Mathematical and Theoretical}, 42(50):504009, 2009.

\bibitem{gomez2022role}
Fernando~J G{\'o}mez-Ruiz, David Subires, and Adolfo del Campo.
\newblock Role of boundary conditions in the full counting statistics of topological defects after crossing a continuous phase transition.
\newblock {\em Physical Review B}, 106(13):134302, 2022.

\bibitem{kou2023varying}
Han-Chuan Kou and Peng Li.
\newblock Varying quench dynamics in the transverse ising chain: The kibble-zurek, saturated, and presaturated regimes.
\newblock {\em Physical Review B}, 108(21):214307, 2023.

\end{thebibliography}

\end{document}